\documentclass[a4paper,11pt]{article}
\pdfoutput=1 

\usepackage{jheppub} 

\usepackage{fancyhdr}
\usepackage{graphicx}

\usepackage{cancel}
\usepackage{braket}

\newcommand{\gl}{\lambda}

\newcommand{\be}{\begin{equation}}
\newcommand{\ee}{\end{equation}}
\newcommand{\bea}{\begin{eqnarray}}

\newcommand{\eea}{\end{eqnarray}}

\newcommand{\Rmnum}[1]{\expandafter\@slowromancap\romannumeral #1@}

\begin{document} 


 \title{\boldmath Large $A_{t}$ Without the Desert}
 
 \begin{flushright}
WITS-CTP-134\\
 LYCEN-2014-07
 \end{flushright}

\author[\spadesuit]{Ammar Abdalgabar,}
\author[\spadesuit]{Alan S. Cornell,}
\author[\clubsuit]{Aldo Deandrea\footnote{also Institut Universitaire de France, 103 boulevard Saint-Michel, 75005 Paris, France}}
\author[\spadesuit]{and Moritz McGarrie}

\affiliation[\spadesuit]{National Institute for Theoretical Physics; School of Physics, University of the Witwatersrand,
Wits 2050, South Africa}
\affiliation[\clubsuit]{Universit\'e de Lyon, F-69622 Lyon, France; Universit\'e Lyon 1, CNRS/IN2P3, UMR5822 IPNL, F-69622 Villeurbanne Cedex, France}

\emailAdd{aabdalgabar@gmail.com}
\emailAdd{alan.cornell@wits.ac.za}
\emailAdd{deandrea@ipnl.in2p3.fr}
\emailAdd{moritz.mcgarrie@wits.ac.za}


\abstract{Even if the unification and supersymmetry breaking scales are around $10^6$ to $10^{9}$ TeV,  a large $A_t$ coupling may be entirely generated at low energies through RGE evolution in the 5D MSSM.  Independent of the precise details of supersymmetry breaking, we take advantage of power law running in five dimensions and a compactification scale in the $10-10^3$ TeV range to show how the gluino mass may drive a large enough $A_t$ to achieve the required $125.5$ GeV Higgs mass. This also  allows for sub-TeV stops, possibly observable at the LHC, and preserving GUT unification, thereby resulting in improved naturalness properties with respect to the four dimensional MSSM. The results apply also to models of ``split families'' in which the first and second generation matter fields are in the bulk and the third is on the boundary, which may assist in the generation of light stops whilst satisfying collider constraints on the first two generations of squarks. }


\maketitle
\flushbottom


\section{Introduction} \label{sec:intro}

\par The discovery of a  scalar particle of mass  $m_h \sim 125.5$ GeV \cite{Aad:2012tfa,Chatrchyan:2012ufa}, consistent with the Standard Model (SM) Higgs boson, in the context of the Minimal Supersymmetric SM (MSSM), motivates considering models of supersymmetry breaking in which stops masses are heavy, of the order of $10$ TeV or greater, or models in which a sufficiently large $A_t$ can be generated at low scales.  In most models of supersymmetry breaking, choosing heavy stops results in the entire coloured sparticle spectrum becoming rather heavy, beyond the reach of the LHC, and is consequently phenomenologically less interesting\footnote{Some recent interesting alternatives may be found in \cite{Bharucha:2013ela,Brummer:2013upa,Abel:2014fka}.}. The second possibility, of large $A_t$, allows for light stops perhaps below $1$ TeV, which is allowed by current collider bounds \cite{ATLAS3rd,CMS3rd} and is aesthetically preferred as it greatly reduces the required fine tuning of the Higgs mass from $\delta m_{H_u}^2$.

\par Models of supersymmetry breaking with a  large $A_t$ at the electroweak scale are usually considered rather difficult to obtain however. For example, in a generic supergravity mediated scenario, one should expect all trilinear soft breaking terms, $A_{u/d/e}(i,j)$, to be of the same order, such that a model in which $A_u(3,3)=A_t$ is sufficiently large is already excluded by flavour constraints on the other off-diagonal elements. Additional {\it ad hoc} symmetries are then required without motivation, to reduce the soft breaking terms to the diagonal elements only. Equally, in minimal gauge mediated supersymmetry breaking (mGMSB) trilinear terms such as $A_t$ are vanishing at the supersymmetry breaking scale $M$, and a large $A_t$ can only be  generated via a rather long period of renormalisation group (RG) evolution. This requires the supersymmetry breaking scale to be very  high, which is also detrimental to the naturalness of the theory. 

\par A purely radiatively generated $A_t$ does, however, have some positive features: the relative hierarchy of Yukawas and the large size of the top Yukawa, $Y_t$, allows for a hierarchy amongst the trilinear soft breaking terms, in which $A_t$ is driven through RG equations (RGEs) almost entirely from the gluino mass $M_3$, where such a  hierarchy between trilinear breaking terms can naturally satisfy flavour changing neutral current (FCNC) constraints.    It is therefore worthwhile to consider extensions of the MSSM that may accelerate the RGE evolution of $Y_t$ or $A_t$ or both.

\par In this paper we will show that a five dimensional (5D) MSSM with compactification scale of $O(10-10^3)$ TeV\footnote{As in our model the Kaluza-Klein mode of the bulk $U(1)$ supplies a $Z'$, collider exclusions  set a lower bound on the compactification scale to be a O(5) TeV.}, and correspondingly a low unification scale of $10^9$ TeV or lower can naturally, through power law running \cite{Taylor:1988vt},  achieve a large $A_t$ at low scales.  The largeness of $A_t$ is driven by the size of the gluino mass $M_3$, which is necessary to be above collider bounds, but is largely independent of how supersymmetry is broken. We simply assume that  $A_t(M_{GUT})\sim 0$ and is entirely generated through renormalisation.  In addition we have explored the  case when all three generations are on the boundary and the ``split families'' case when the 3rd generation of matter multiplets is on the boundary and the first two are in the bulk.  Our results hold similarly for both case, but the second may be more favourable to generate a hierarchy of soft masses ${m^2_{(Q,U,D)}}_3\ll {m^2_{(Q,U,D)}}_{1,2}$, which should be more natural and phenomenologically more interesting as stops can then be much lighter, and within reach of the LHC.

\par The paper is structured as follows: in section \ref{LargeAt} we describe the setup and explore the RGEs of a number of parameters from the unification scale to the electroweak scale, in particular focusing on achieving a large $A_t$ parameter. In  section \ref{lightstops} we use the achieved values of $A_t$ in our models to estimate the necessary size of the lightest stop mass, to obtain the currently observed Higgs mass, in particular emphasing that the 5D MSSM allows for sub-TeV stops due to the sizable $A_t$. We discuss in section \ref{models} some different scenarios for supersymmetry breaking.  In section  \ref{Discussion} we discuss our results and how this work may be extended. Appendix \ref{appendix1} supplies our conventions for the 5D MSSM and in appendix \ref{RGES5D} we supply the full one-loop 5D RGE's for all supersymmetric and soft term parameters of our model, which is also an important calculational result of this work. 

\begin{figure}[!thb]
\begin{center}
\includegraphics[width=7cm,angle=0]{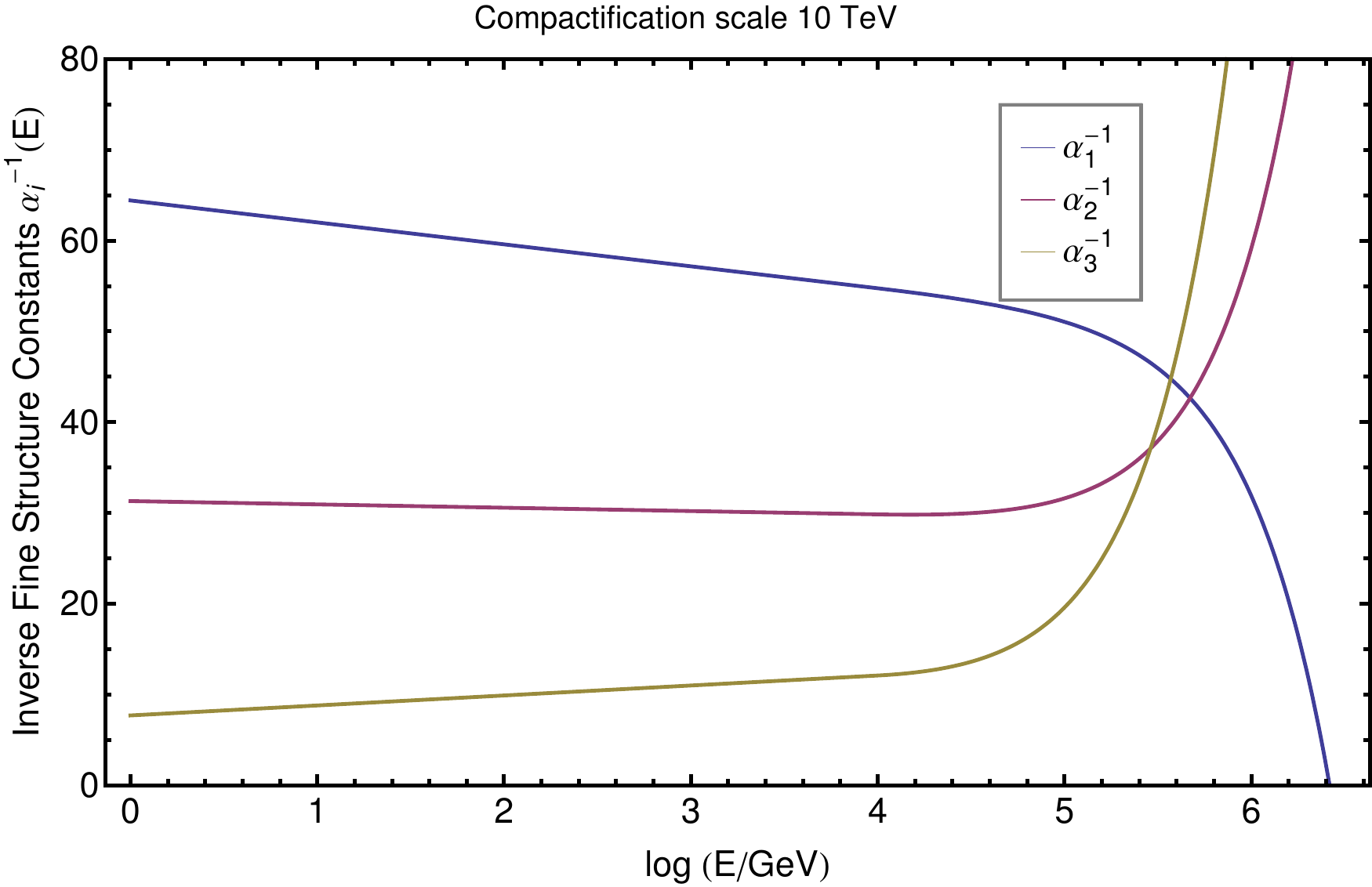}\qquad
\includegraphics[width=7cm,angle=0]{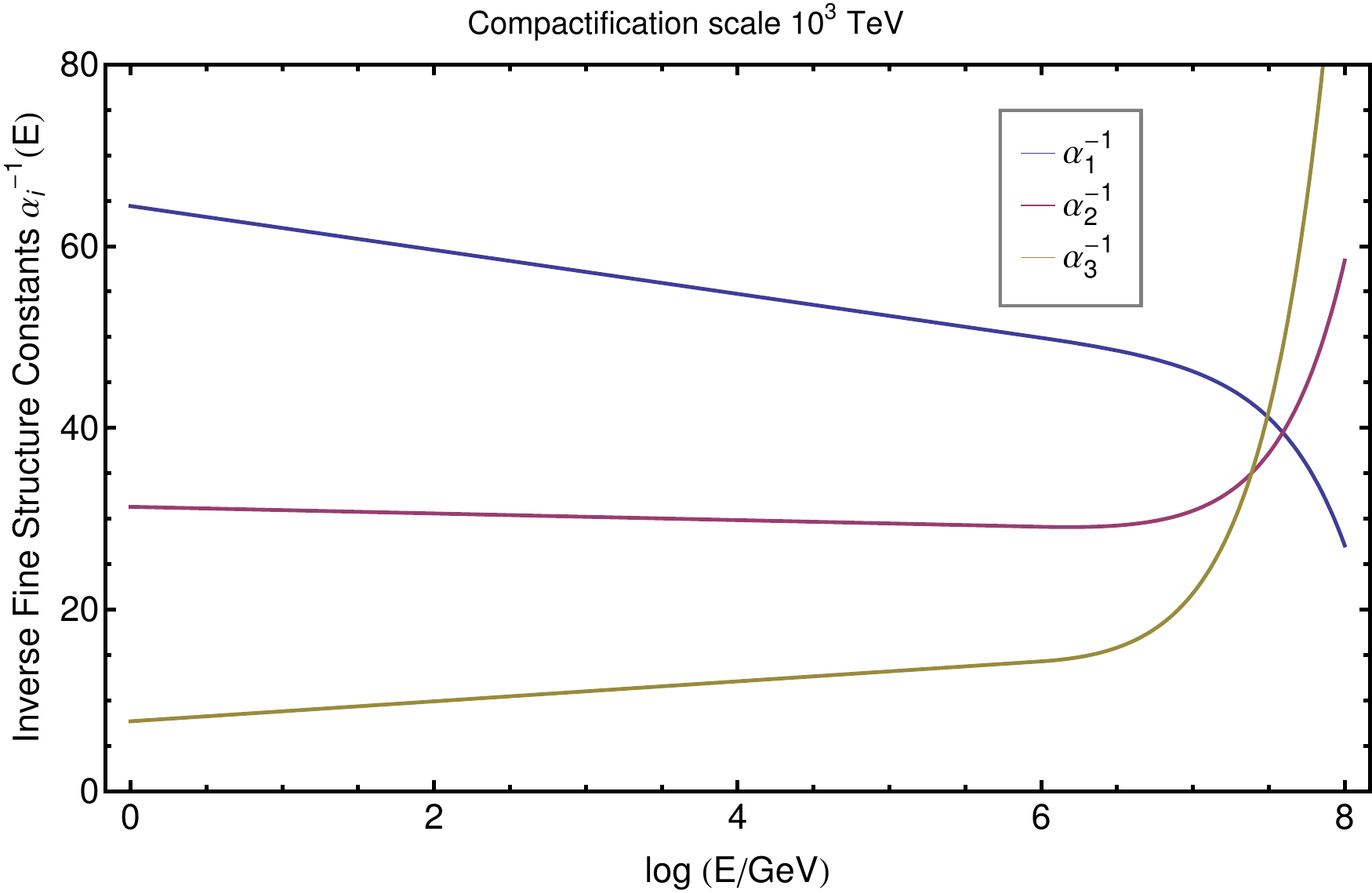}\qquad
\includegraphics[width=7cm,angle=0]{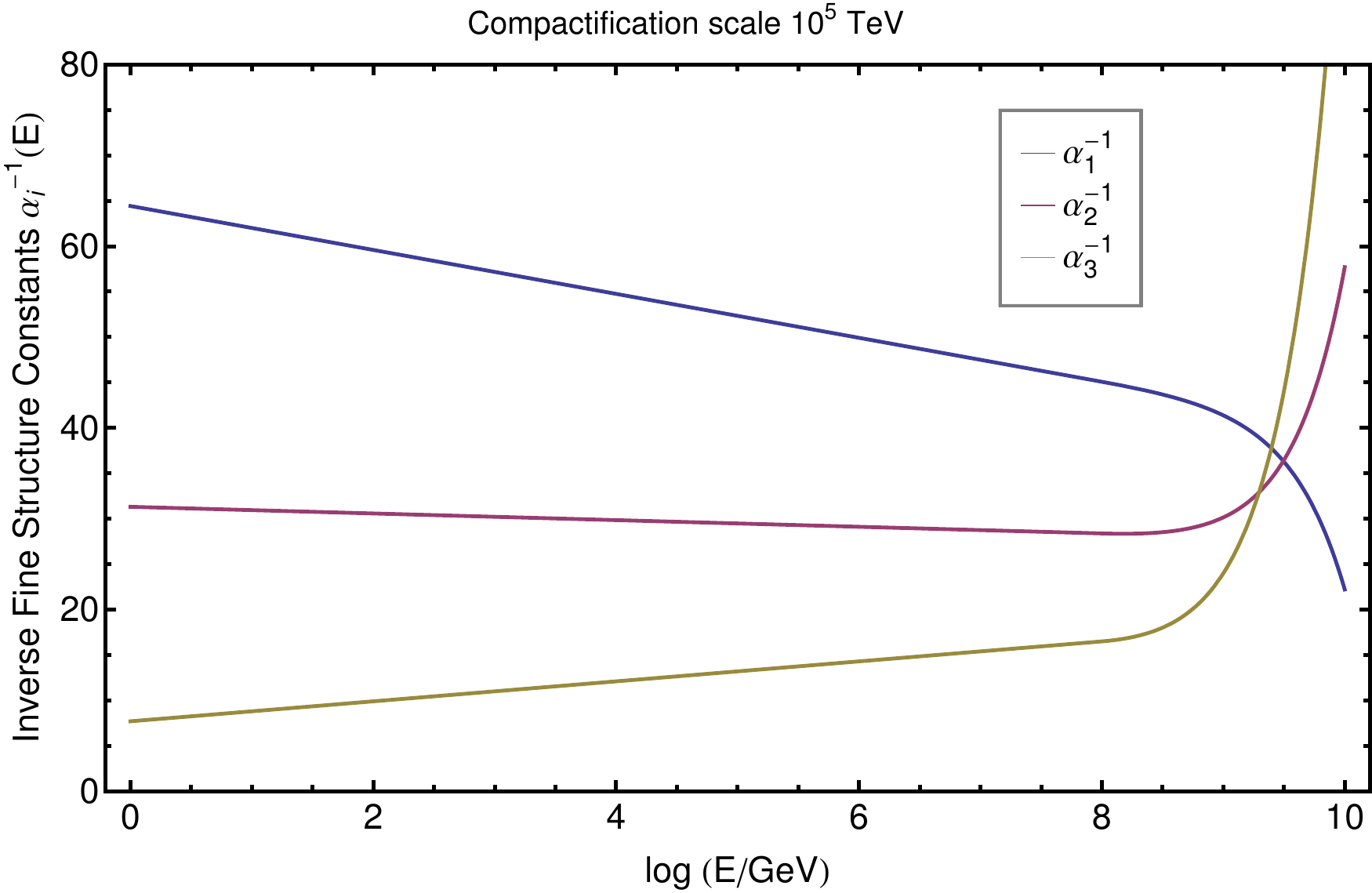}\qquad
\includegraphics[width=7cm,angle=0]{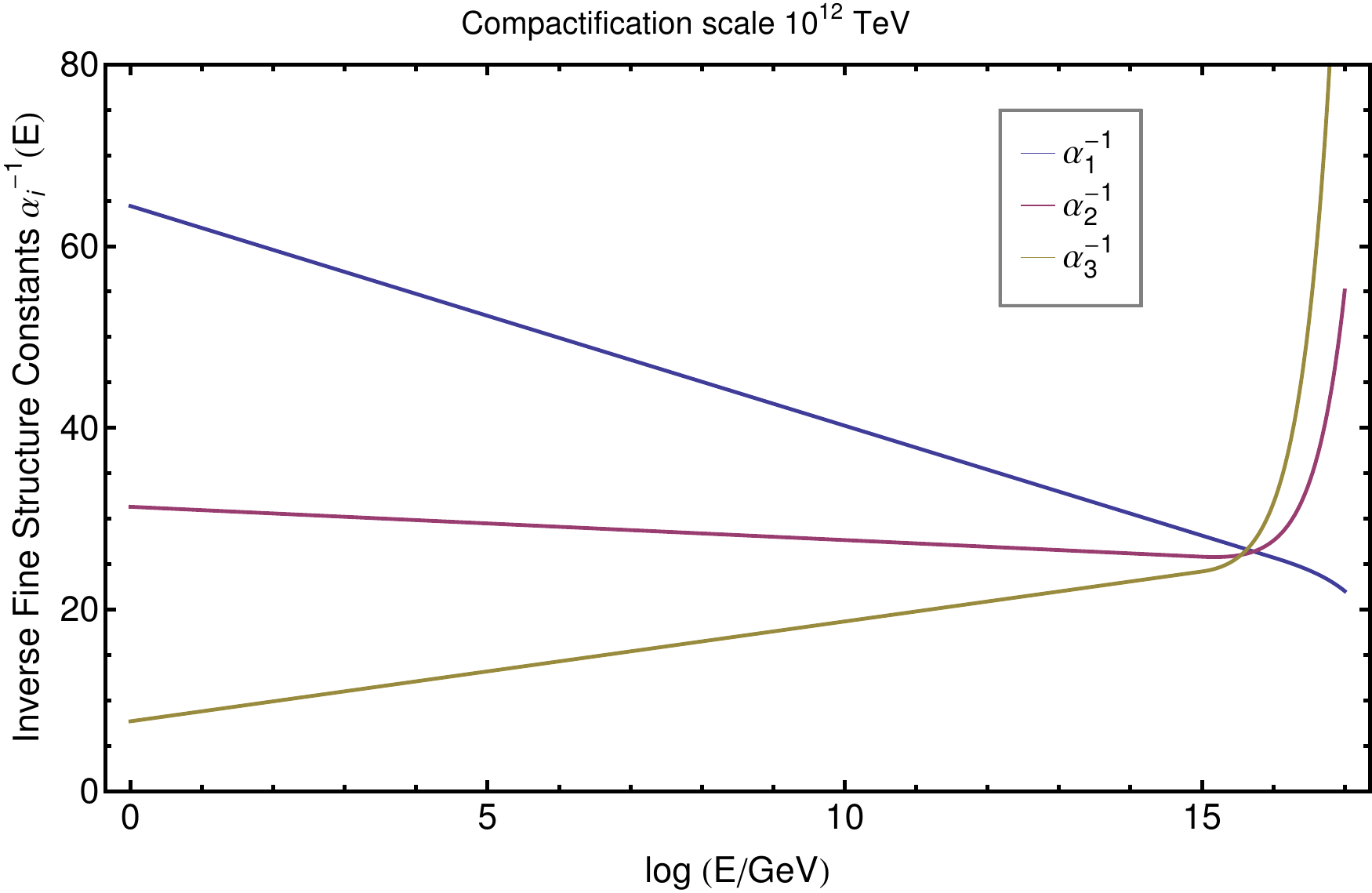}
\caption{{\it Running of the inverse fine structure constants $\alpha^{-1}(E)$, for three different values of the compactification scales 10 TeV (top left panel), $10^3$ TeV (top right), $10^5$ TeV (bottom left) and $10^{12}$ TeV (bottom right), with $M_3$ of 1.7 TeV, as a function of  log(E/GeV).}} 
\label{fig:alphas5D}
\end{center}
\end{figure}


\section{Generating large $A_{t}$ in the 5D MSSM}\label{LargeAt}

\par In this section we describe the details and the setup of our model, we describe our parameterisation of the UV boundary conditions such as the  supersymmetry breaking and the electroweak boundary conditions. We then discuss our results for the evolution of various parameters of our model.

\begin{figure}[t!]
\begin{center}
\includegraphics[width=7cm,angle=0]{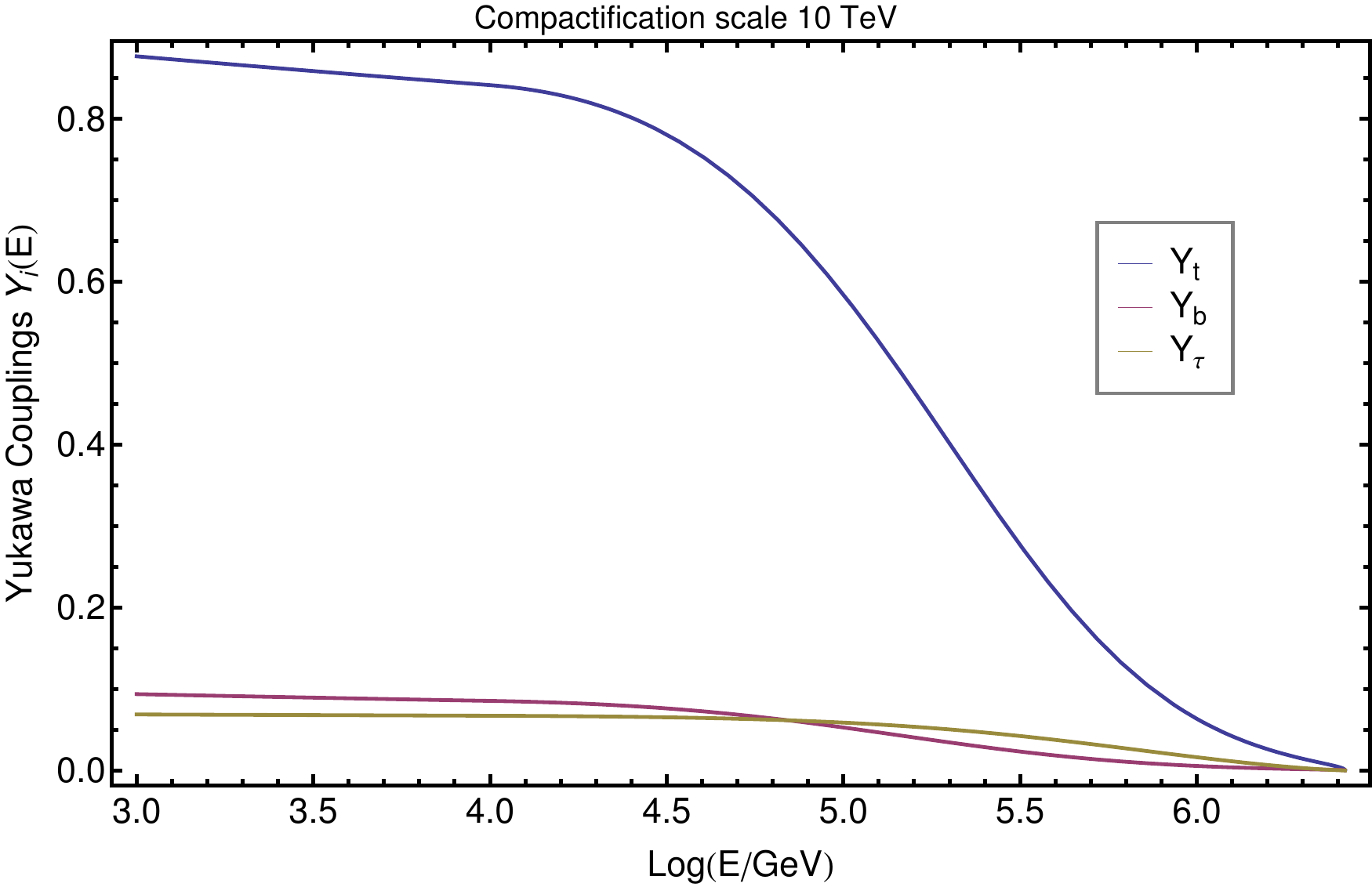}\qquad
\includegraphics[width=7cm,angle=0]{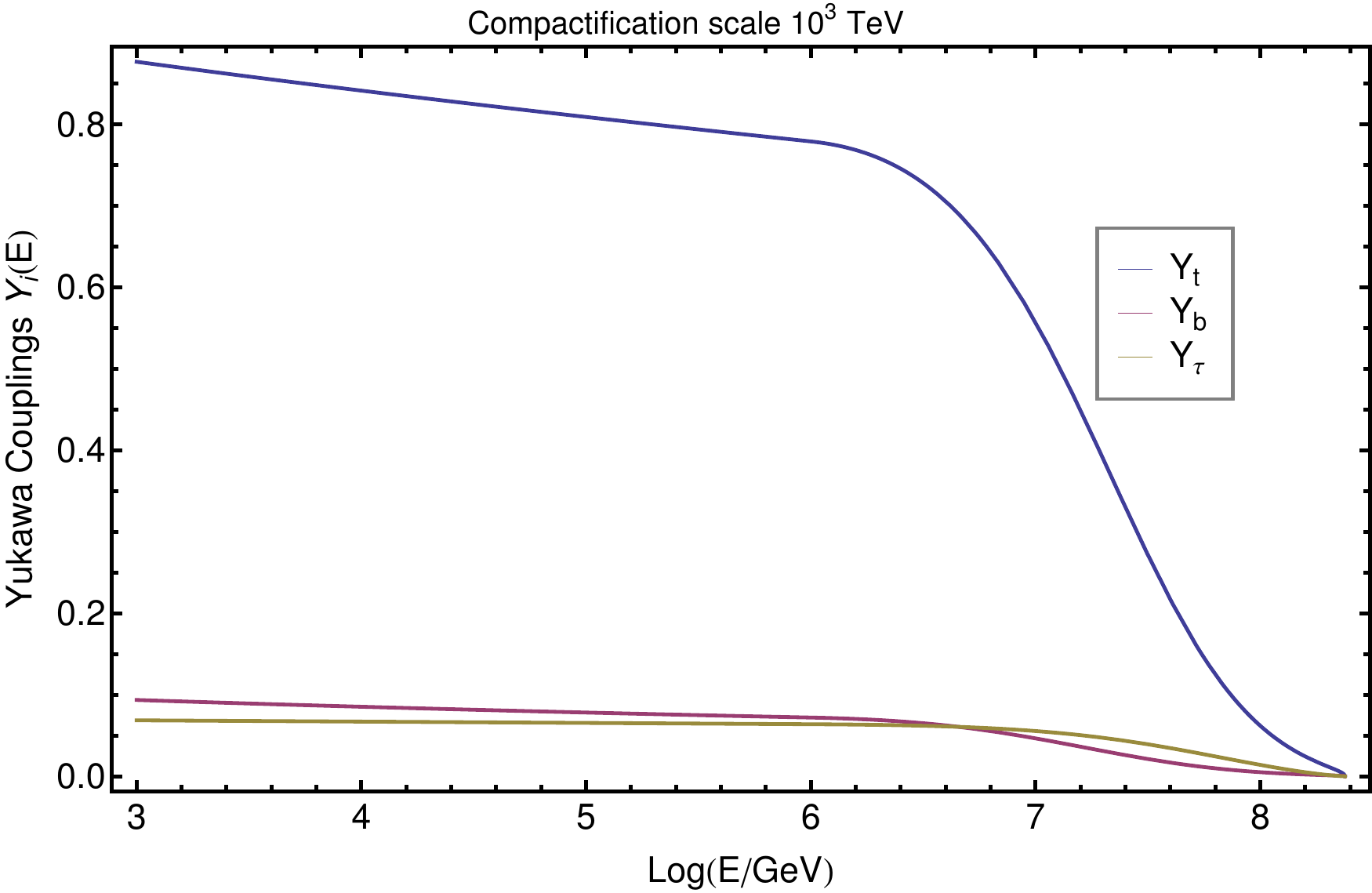}\qquad
\includegraphics[width=7cm,angle=0]{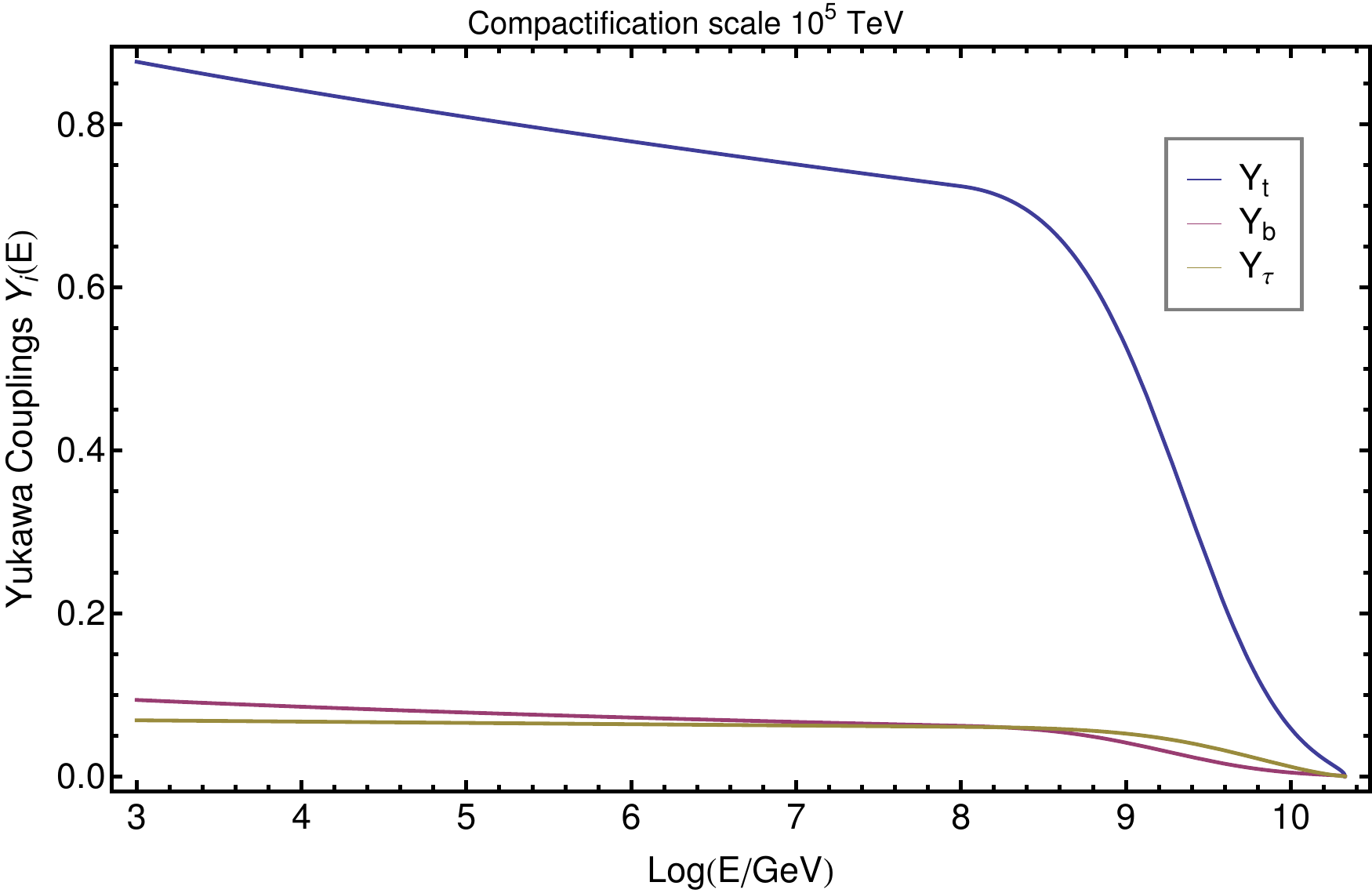}\qquad
\includegraphics[width=7cm,angle=0]{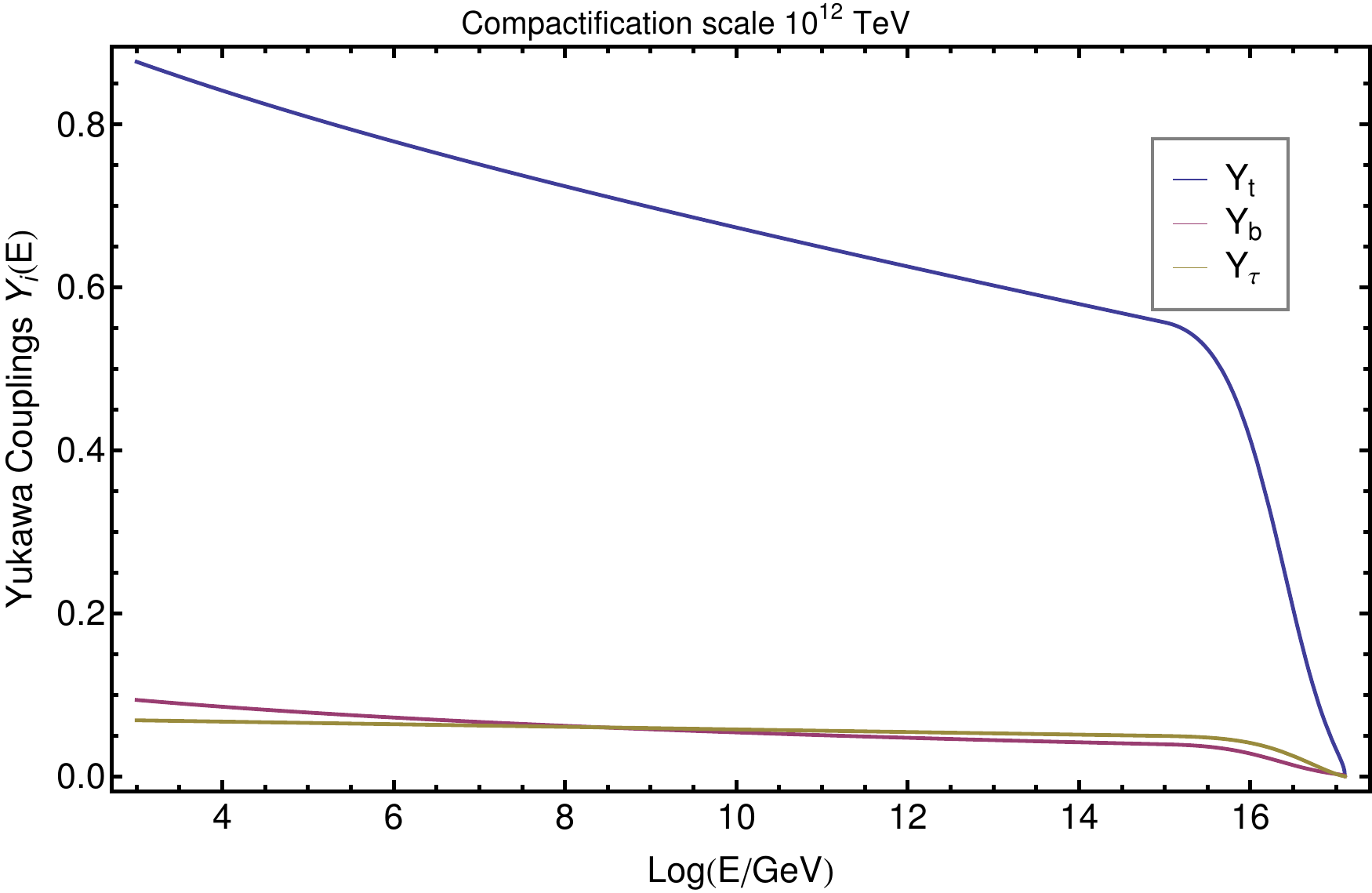}
\caption{{\it Running of Yukawa couplings $Y_i$,  for three different values of the compactification scales: 10 TeV (top left panel), $10^3$ TeV (top right), $10^5$ TeV (bottom left) and $10^{12}$ TeV (bottom right), with $M_3[10^3]$ of 1.7 TeV, as a function of  log(E/GeV).}} 
\label{fig:Yuk5D}
\end{center}
\end{figure}

\subsection{The setup}

\par We define the 5D MSSM to be a field theory on a four dimensional space-time, times an interval of length $R$ in which the SM $SU(3)_c\times SU(2)_L\times U(1)_Y$ gauge fields and the  Higgses ($H_u,H_d$) propagate into the fifth dimension. As a result these fields will have Kaluza-Klein modes which contribute to the RGEs at $Q> 1/R$ and additional matter associated to five dimensional $\mathcal{N}=1$ super Yang-Mills. Different possibilities of localisation for the matter fields can be studied, however we shall consider first the  limiting case with SM matter fields restricted to the $y=0$ brane, and we supply the RGEs for this scenario in appendix \ref{RGES5D}. Therefore there will be no additional Kaluza-Klein contributions of these matter fields to the RGEs.  In a specific setup, only the third family is restricted to the brane, while the light generations are allowed to propagate in the bulk.  Note however that from the point of view of numerical results this case is not much different from restricting all the three generations to the brane, as the only large effects in the renormalisation group evolution are due to the third family coefficients, while the first two generations play only a minor role. Even if in the following we will explicitly discuss the case of all three fermion families restricted to the brane, we have checked numerically that restricting to the brane only the third family does not qualitatively change our conclusions. Note also that five dimensional super Yang-Mills have additional matter fields, such as colour adjoint chiral superfields \cite{Bhattacharyya:2010rm,Bhattacharyya:2012qj}, compared to its four dimensional counterparts and these can influence the RGEs.

\begin{figure}[!th]
\begin{center}
\includegraphics[width=7cm,angle=0]{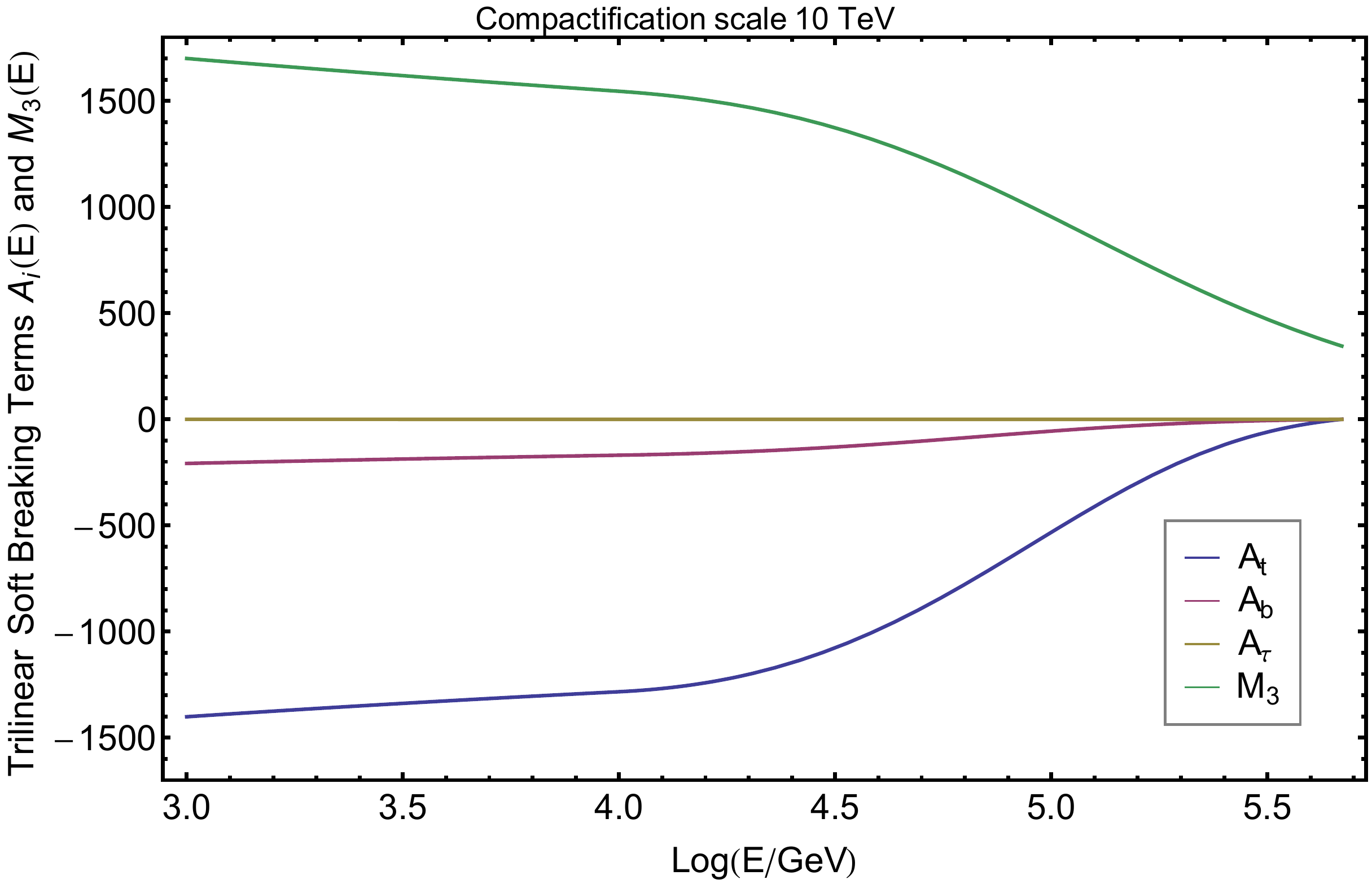}\qquad
\includegraphics[width=7cm,angle=0]{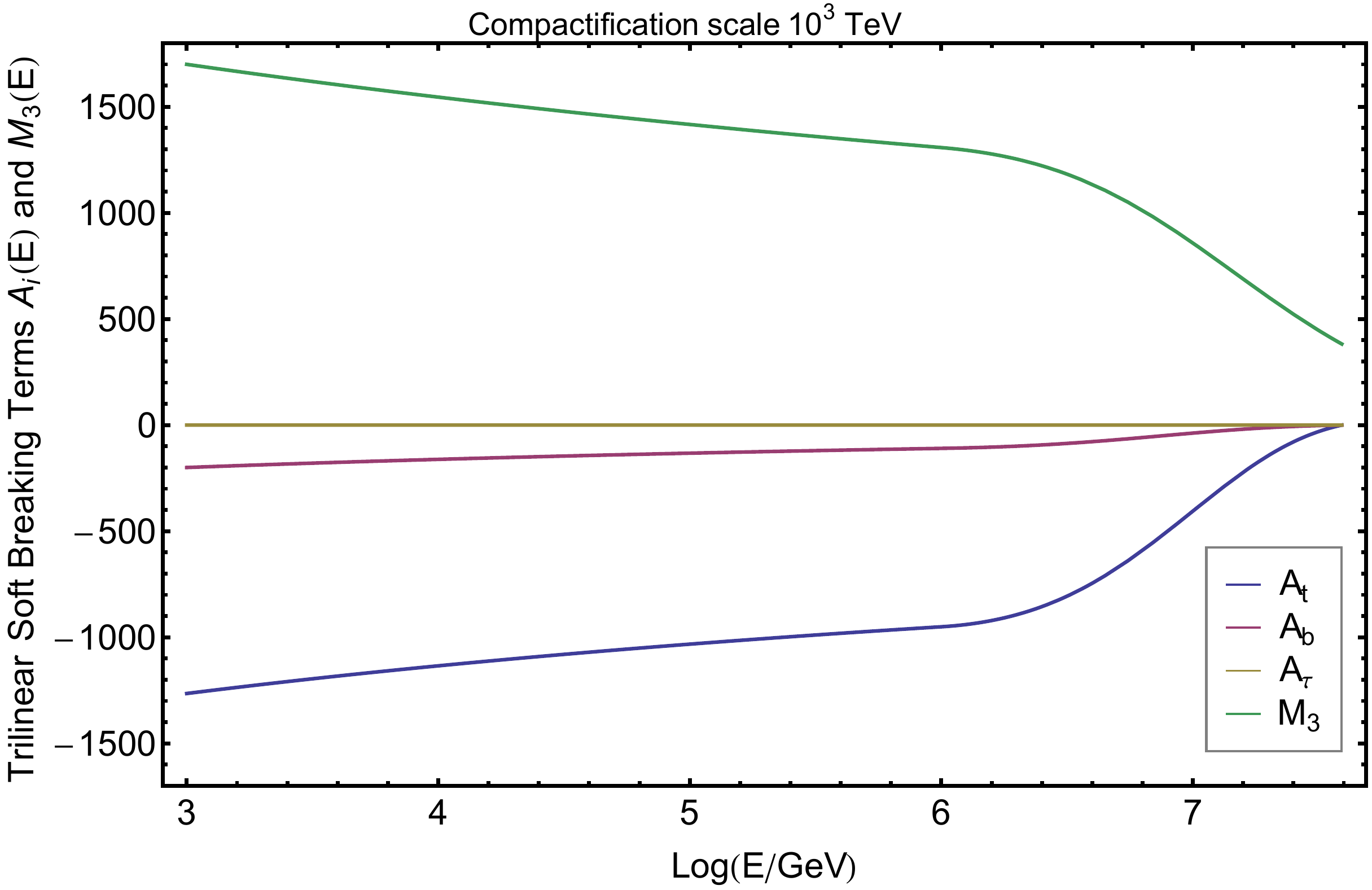}\qquad
\includegraphics[width=7cm,angle=0]{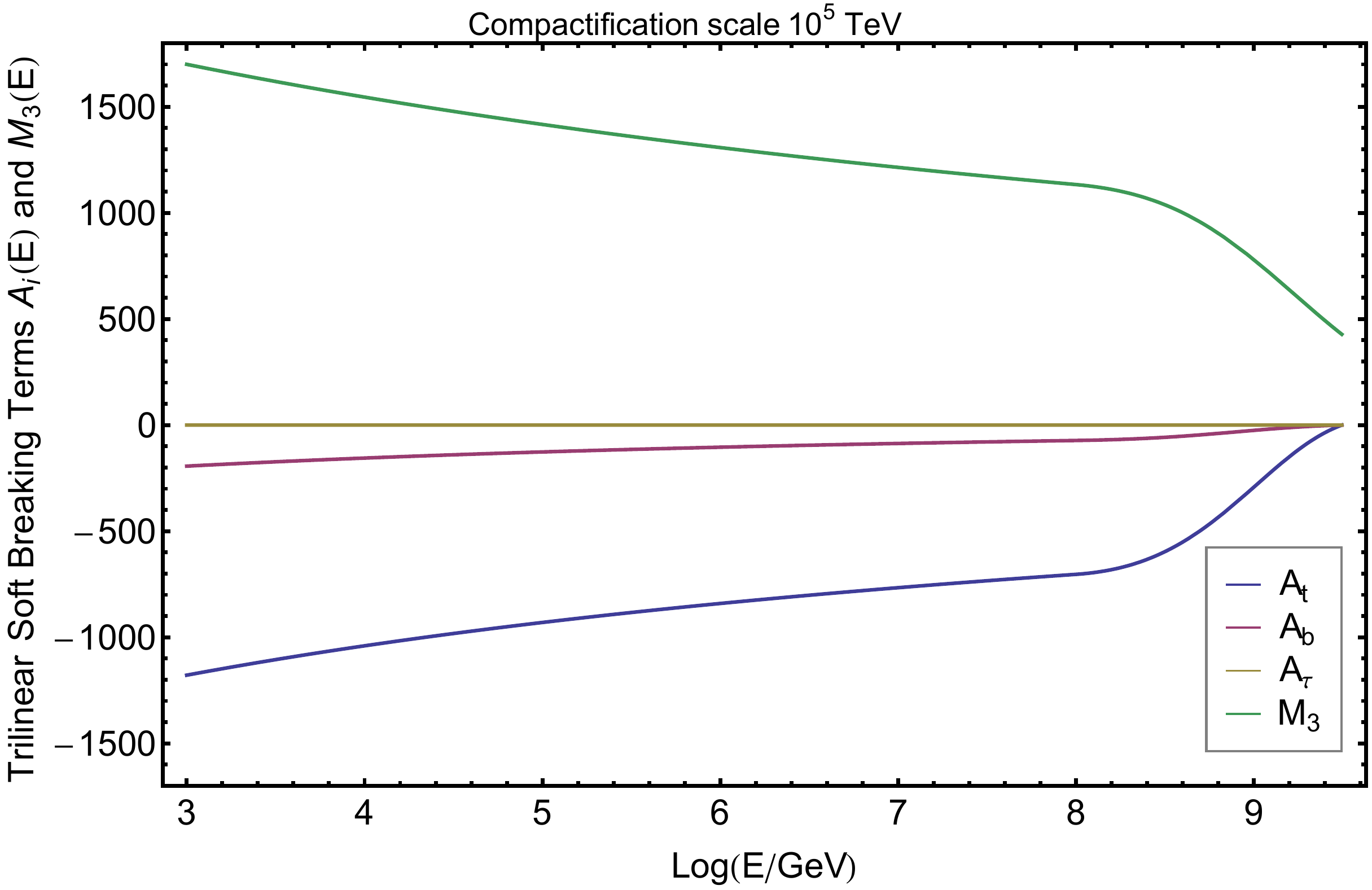}\qquad
\includegraphics[width=7cm,angle=0]{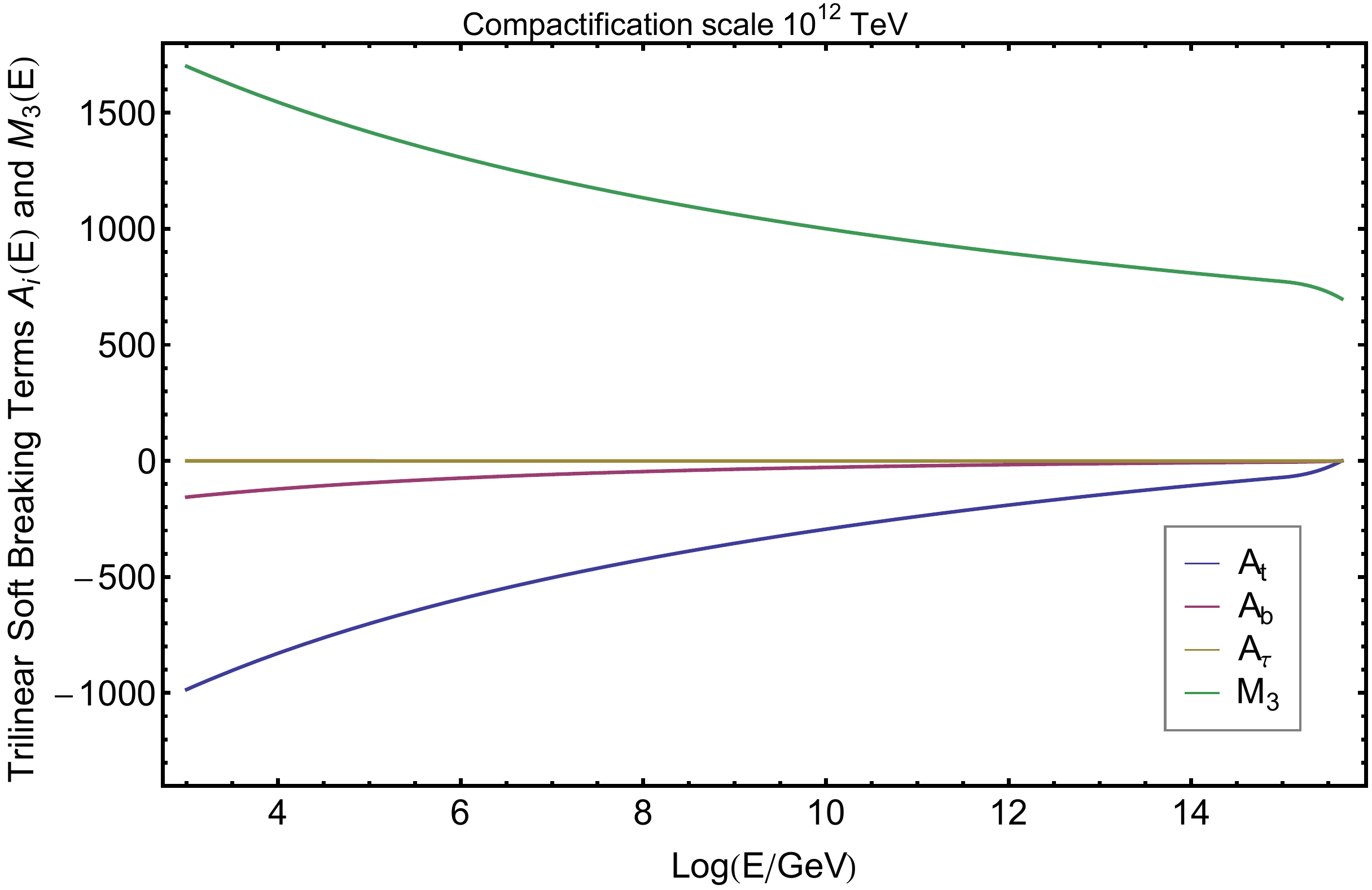}
\caption{{\it Running of trilinear soft terms $A_i(3,3)(E)$, for three different values of the compactification scales 10 TeV (top left panel), $10^3$ TeV (top right), $10^5$ TeV (bottom left) and $10^{12}$ TeV (bottom right), with $M_3[10^3]$ of 1.7 TeV, as a function of   log(E/GeV).}}
\label{fig:trilinearcompact}
\end{center}
\end{figure}

\par Regarding the breaking of supersymmetry, whilst gauge mediation is favoured (and some recent work on gauge mediated supersymmetry breaking in a five dimensional context may be found in \cite{McGarrie:2010kh,McGarrie:2010yk,McGarrie:2011av,McGarrie:2011dc,McGarrie:2012fi,McGarrie:2012ks,McGarrie:2013hca}), ultimately the universality of squark massses in GMSB mean that even though the gaugino mediated limit \cite{Mirabelli:1997aj,Chacko:1999mi,Schmaltz:2000gy,Schmaltz:2000ei} might allow for light squarks (and 5D RGE evolution allows for a large $A_t$ and the observed Higgs mass), the collider bounds on first and second generation squarks \cite{ATLASsqgl,CMSsqgl}, in the supra-TeV range would apply also to the $3^{rd}$ generation squarks, i.e. the stops, which as discussed before, is both phenomenologically less interesting and unnatural.  Therefore we wish for some other description of supersymmetry breaking that may allow for stops to be lighter than their first and second generation counterparts, such as in \cite{Brummer:2013upa,Abel:2014fka}. In this paper we will therefore be rather agnostic about the precise details of how supersymmetry is broken and as a result also our conclusions will apply quite generally. We do however make some minimal specifications:
\begin{itemize}
\item We take as inputs the Yukawa and gauge couplings at the SUSY scale, $1$ TeV.
\item We will assume supersymmetry breaking occurs at the unification scale, which is found by finding the scale at which $g_1=g_2$, which is lowered compared to the 4D MSSM, by the effects of the compactification.
\item We specify the value of the gluino mass, $M_3$ at  $1$ TeV.
\item We take the trilinear soft breaking terms, $A_{u/d/e}$, to vanish at the unification scale.
\end{itemize}
Our procedure is to solve the combined set of differential equations numerically using the above conditions, taking the ``third family'' approximation in which we only evolve third generation RGEs, although the full RGEs are supplied in appendix \ref{RGES5D}. This approximation is quite standard and is due to the relative smallness of the other Yukawa couplings (at least one order of magnitude) compared to those of the third generation and as a result the other A-term values are also very small.  We further specified some parameters such as $\mu$, $B_{\mu}$ and the value of the sfermion masses ($\sim 1$ TeV) so as to allow for the RGEs to be solved, but these do not affect the overall result.  We solved the differential equations between $Q_{min}=10^3$ GeV and $Q_{max}$, which was typically only one order larger than the unification scale, for each scenario explored. The details of the RGEs and how the Kaluza-Klein summation is accounted for is discussed in appendix \ref{RGES5D}.  

\par An interesting feature of the 5D MSSM is the approximate unification of gauge couplings \cite{Castano:1993ri,Dienes:1998vh,Dienes:1998vg,Pomarol:2000hp,Nomura:2006fz}, which is here calculated to one-loop and presented in figure \ref{fig:alphas5D} for various compactification scales.   The key feature of figure \ref{fig:alphas5D} is that with a larger compactification radius the unification scale can be significantly lowered, lowering the desert of scales between the electroweak scale and unification.   In this paper we will take the unification scale to be the scale of supersymmetry breaking such that a lower supersymmetry breaking scale will also assist in improving the naturalness of each model, as we shall see later.

\begin{figure}[!th]
\begin{center}
\includegraphics[width=7cm,angle=0]{./figures/Trilinear10TeV1700M_3new.pdf}\qquad
\includegraphics[width=7cm,angle=0]{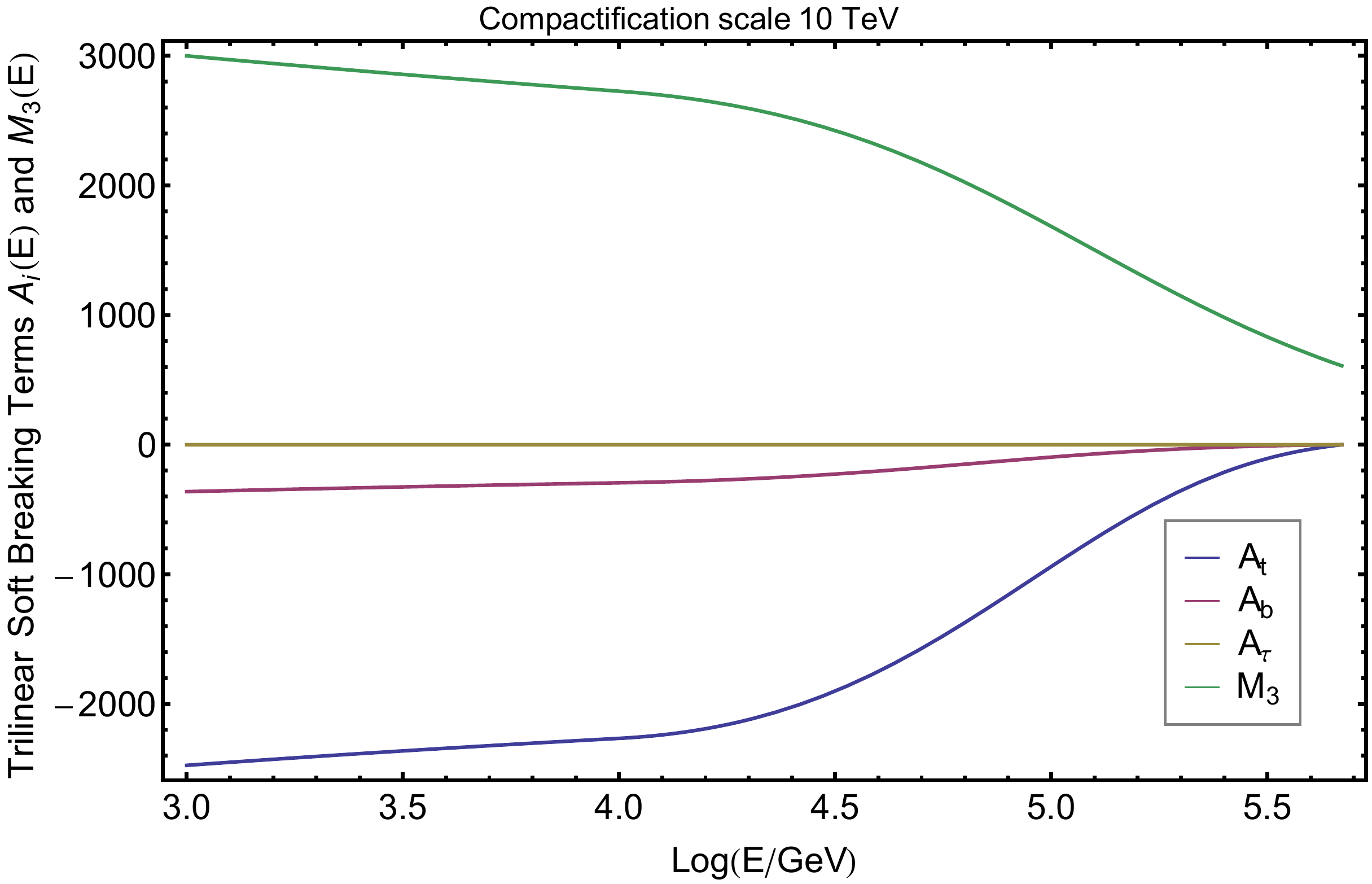}\qquad
\includegraphics[width=7cm,angle=0]{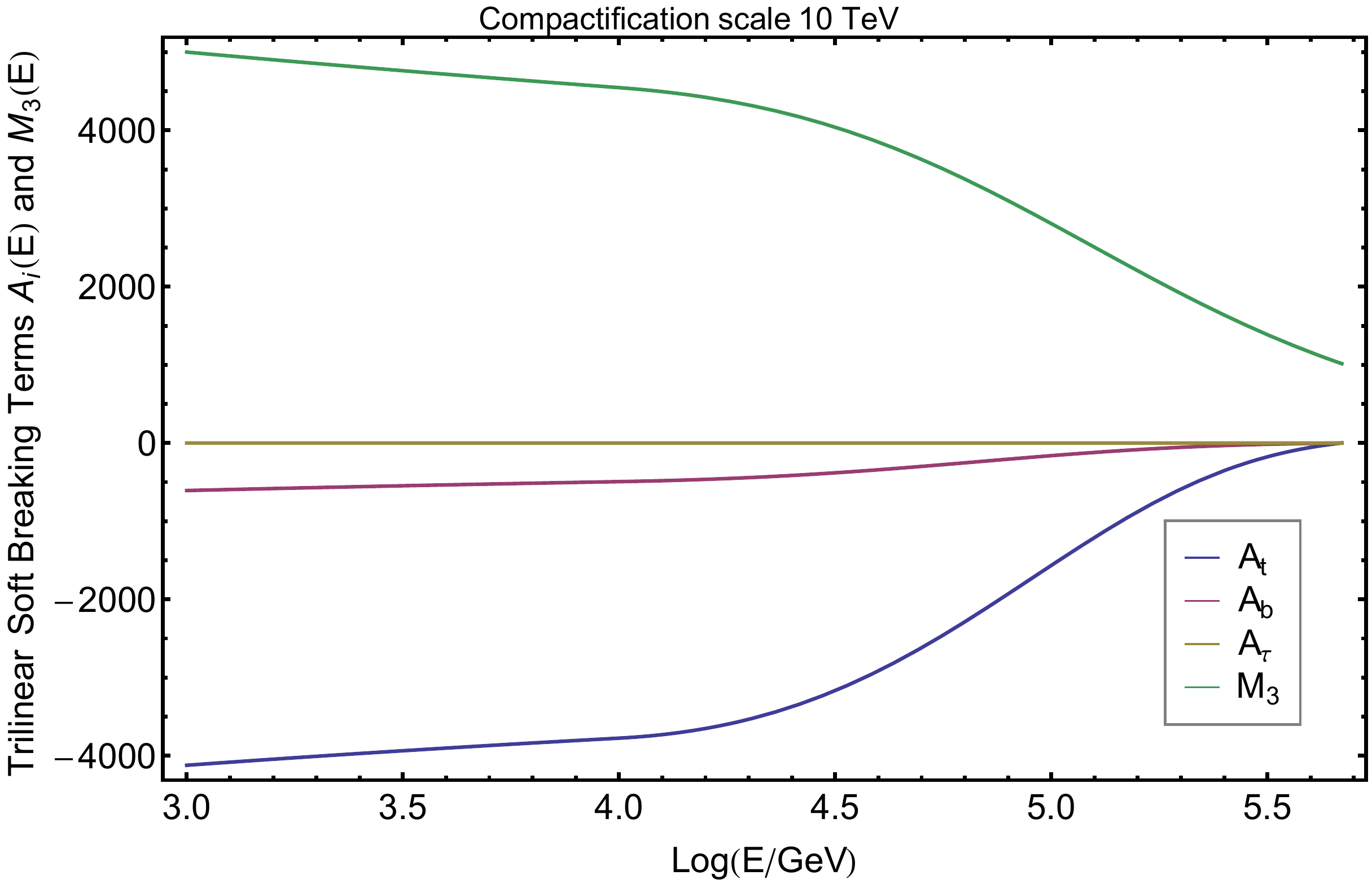}
\caption{{\it Running of trilinear soft terms $A_i(3,3)(E)$, for three different values of gluino masses, $M_3$: 1.7 TeV (top left panel), 3 TeV (top right panel) and 5 TeV (bottom panel), with $R^{-1}$ of 10 TeV, as a function of  log(E/GeV).}}
\label{fig:trilineargluino}
\end{center}
\end{figure}

\par We also specify the Yukawa coupling RGE \cite{Cornell:2011fw,Cornell:2012qf,Cornell:2012uw,Abdalgabar:2013xsa,Abdalgabar:2013oja} boundary conditions at $1$ TeV, which interestingly appear to vanish when evolved to the unification scale as shown in figure \ref{fig:Yuk5D}.

\par Let us now focus on the evolution of the $A_t$ terms.  As mentioned before, we fix a low scale value of the gluino mass $M_3$ and set a high scale boundary condition that the $A_i$'s vanish, and then solve the set of equations.  The results are presented in figure \ref{fig:trilinearcompact} for various compactification radii, and then for a fixed radius of $10$ TeV but for varying gluino mass $M_3$ in figure \ref{fig:trilineargluino}.  We see in  figure \ref{fig:trilinearcompact}  that by increasing the compactificaton radius one can increase the size of the trilinear soft breaking term. Figure \ref{fig:trilineargluino} shows that after a reasonable period of RG evolution the $A_t$ mimics the magnitude of the final value of the gluino mass, at $1/R\sim 10$ TeV, such that at low scales $|A_t|\sim M_3$. Therefore, for this compacitification radius an $O(2)$ TeV gluino can generate a reasonably large size $A_t$ at low scales, but with an initially low unification scale.  If we associate the unification scale with the Messenger scale, which is where we assume the A-terms to vanish, in the context of GMSB for example, this suggests that we can still have a low messenger scale of $10^6-10^9$ GeV, for a sufficiently large compactification radius.  Equally we could have a small compactification radius, in which case we would need a very high initial scale of running to obtain similar sized A-terms, which is detrimental to the naturalness of the theory, as pictured figure \ref{fig:trilinearcompact} bottom right panel.  To summarise, we may achieve a large $A_t$ term by exchanging a high initial supersymmetry breaking scale such as in the four dimensional MSSM, for a larger compactification radius and a lower initial supersymmetry breaking scale. Such a scenario has improved naturalness properties and is favourable from this perspective.


\section{Light Stops Without the Desert}\label{lightstops}

\begin{figure}[!thb]
\begin{center}
\includegraphics[width=0.9\textwidth]{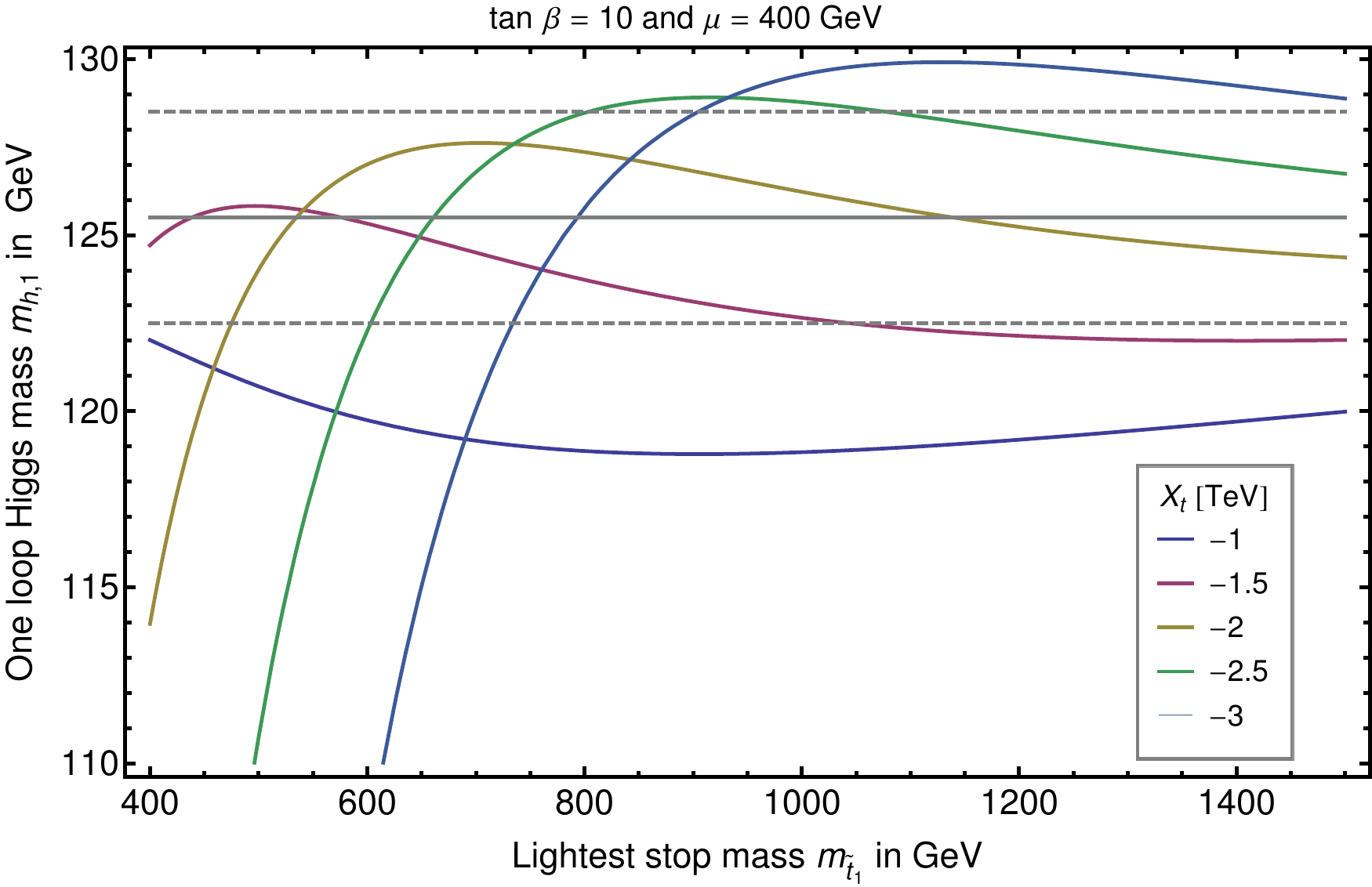}
\caption{{\it A plot of the one loop Higgs mass versus the lightest stop mass for representative values of $X_t=A_t-\mu \cot \beta$, corresponding to those of the 5D MSSM. }} 
\label{fig:HiggsmassforAt}
\end{center}
\end{figure}

\par An important result of obtaining large $A_t$ at low scales is that one may then achieve the correct Higgs mass with a lower stop mass scale. Using the (MSSM) one-loop Higgs mass in the limit $m_{A^0}\gg m_Z$ ~\cite{Ellis:1991zd,Lopez:1991aw,Carena:1995bx,Haber:1996fp,Degrassi:2002fi}  one has
\be
m_{h,1}^2\simeq m_{z}^2\cos^2 2\beta +
\frac{3}{4\pi^2}\frac{m_t^4}{v^2_{ew}}\left[\ln
\frac{M^2_{S}}{m_t^2}+\frac{X^2_t}{M_S^2}\left(1-\frac{X^2_t}{12M_S^2}
\right)\right],\label{eq:higgsmass}
\ee
where $v_{ew}$ is the electroweak Higgs vev, $X_t=A_t -\mu \cot \beta$ and $M^2_S=m_{\tilde{t}_1}m_{\tilde{t}_2}$.   Fixing $m_{h,1}=125.5$ GeV, $m_Z=91$ GeV, $\mu=200$ for $\tan \beta =10$ we can see in figure \ref{fig:HiggsmassforAt} that for representative values of $A_t$ achievable in the 5D MSSM, one may easily accommodate the lightest stop mass in the sub-TeV range.

\par
\begin{figure}[h!]
\begin{center}
\includegraphics[width=7.5cm,angle=0]{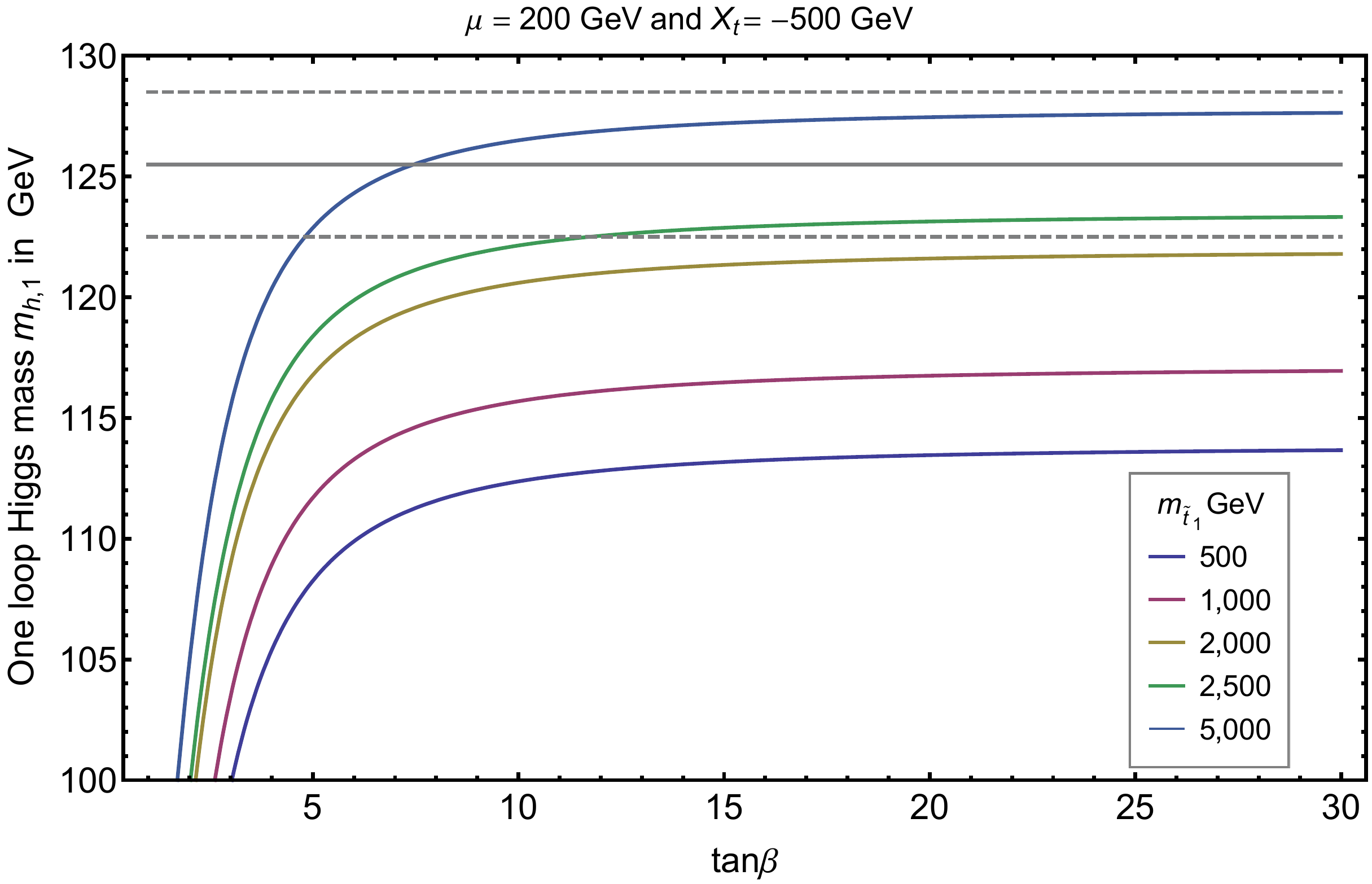}
\includegraphics[width=7.5cm,angle=0]{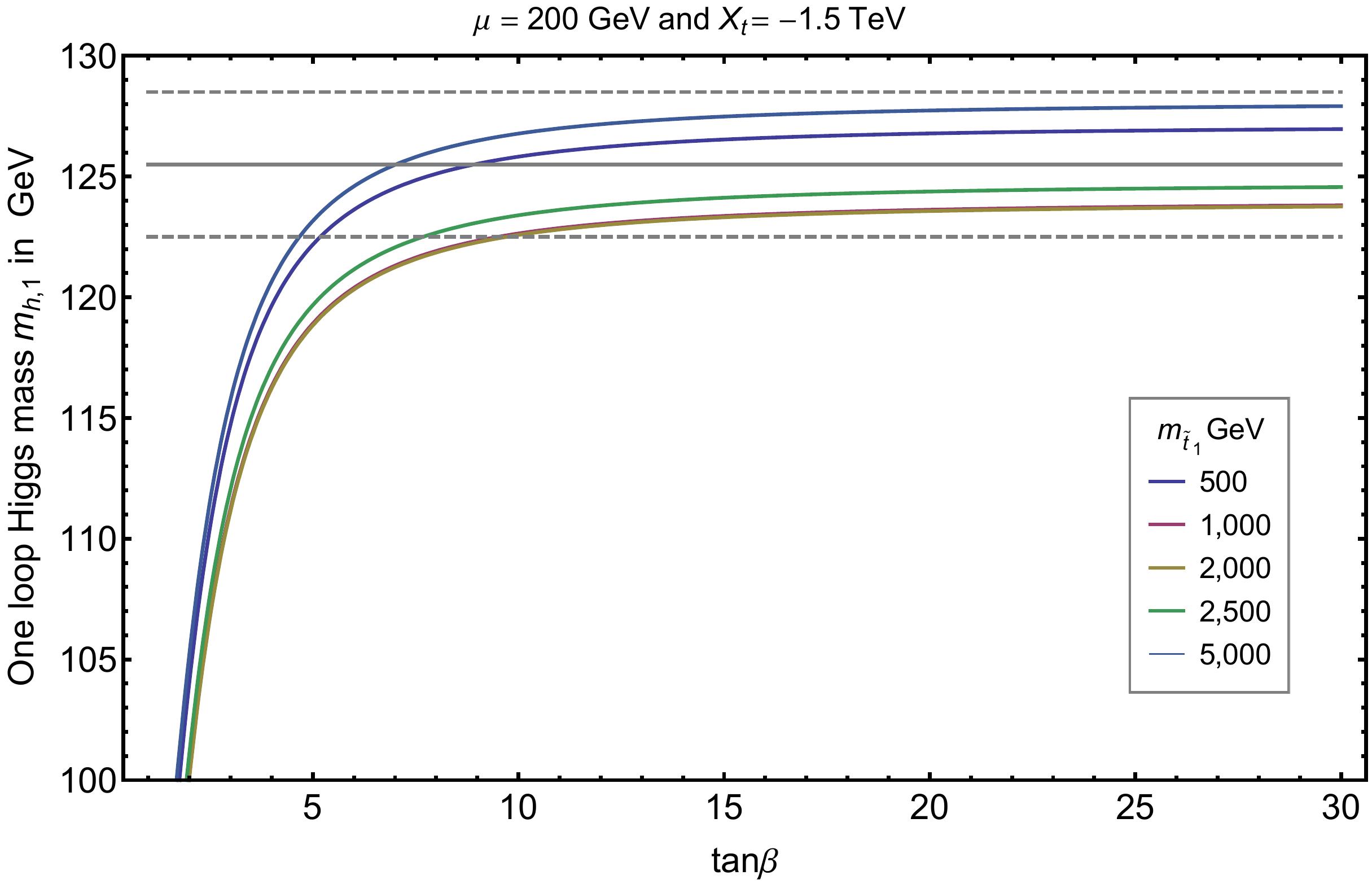}
\caption{{\it A plot of the one loop Higgs mass versus $\tan \beta$ for different values of the stop mass, for  $X_t=A_t-\mu \cot \beta$ of $-500$ GeV (left panel) and $-1.5$ TeV (right panel). }} 
\label{fig:Higgsmassfortandbeta}
\end{center}
\end{figure}

 Let us also discuss the model's dependence on the value of $\tan \beta$ as pictured in figure \ref{fig:Higgsmassfortandbeta}.  The precise value of $\tan \beta$ will depend greatly on how $\mu$ and $B_{\mu}$ are addressed in the context of supersymmetry breaking and hence the solution of the vacuum tadpole equations, but regardless of this, for values of $\tan \beta >10$ the functions are approximately flat and we expect the value to fall within this interval.  We expect that the $\mu$ term is naturally of the order of the electroweak scale, where in figure \ref{fig:HiggsmassforAt} we took a slightly large $\mu$ value of $400$ GeV and in figure \ref{fig:Higgsmassfortandbeta} we took $200$ GeV, leading typically to light Higgsinos and winos.

\par These models have an  interesting additional naturalness feature: the lowered unification and supersymmetry breaking scale necessary compared to the 4D MSSM, results in a lowered cutoff  to radiative corrections, for example, on stops from the gluino:
\be
\delta m_{\tilde{t}}^2=\frac{2g_3^2}{3\pi^2}M_3^2\log \left(\frac{M_{\cancel{SUSY}}}{M_3}\right).
\ee
If the susy breaking scale can then be kept low enough, this can allow for stops remaining light as well as reduced radiative corrections on the Higgs mass,
\be
\delta m_{H_u}^2=-\frac{3y_t^2}{8\pi^2}(m^2_{Q_3}+m^2_{U_3}+A_t^2)\log \left(\frac{M_{\cancel{SUSY} }}{m_{\tilde{t}}}\right)\,.
\ee
The details will depend on how supersymmetry is parameterised at the SUSY breaking scale and as such will be part of a future study, however it should be clear that an $M_{\cancel{SUSY}}\sim M_{GUT}$ of $10^6$ GeV would fair much better than $10^{16}$ GeV, with regard to radiative corrections to fine tuning.

\section{Compatible models of supersymmetry breaking}\label{models}
\par As the feature of a large $A_t$ term from RG evolution with a small compactification scale is rather generic, we have so far been agnostic about the specific details of how supersymmetry is broken.  There are a number of models of supersymmety breaking that may be compatible with our setup so here we describe them and some additional features of the sparticle spectrum that we can infer.

\subsection{Sequestered super-gravity mediation}
\par Four dimensional super-gravity mediation has a number of issues that need to be overcome.  Firstly the theory is non-renormalisable and as such one-loop calcutions of the soft masses should not be trusted. Even if the resulting soft masses are all set from dimensional analysis arguments, this leads to large FCNCs as all entries in the $A_{u/d/e}(3,3)$ would be of the same order, as discussed in the introduction.  Further one should generically expect large mixings between the K\"ahler potentials of the visible sector and SUSY breaking matter fields, such that soft scalar masses are not flavour universal.

\par Sequestered or brane to brane super-gravity mediation \cite{Rattazzi:2003rj,Gregoire:2004nn,Scrucca:2004cw,Falkowski:2005zv,Diamandis:2013zba}
 overcomes many of these drawbacks: supersymmetry breaking effects are calculable and finite at one-loop.  Mixing of K\"ahler potentials at tree level does not arise due to spatial separation of the visible and hidden sectors.  In this scenario, A-terms would be vanishing at the high scale and our results might then be compatible with this scenario by having purely radiatively induced A-terms.  Sequestered supergravity mediation is therefore a favourable model compatible with our results.

\par Even though we do not specify many details of the setup, we may already make some comments on the sort of spectrum of this scenario:
\begin{itemize}
\item[$\bullet$] The lightest superparticle may be the sneutrino (stau), neutralino (neutral wino, bino or Higgsino), generically.
\item[$\bullet$]  The gravitino mass is given by $M_{3/2}\sim \frac{F}{\sqrt{3}M_{Pl}}$ and may arguably be related to that of the gluino mass, $M_3= -3\frac{g_3^2}{16\pi^2} m_{3/2}$, which we took to be just above current exclusion, $1.7$ TeV.  
\end{itemize}
Any physical effect due to ``anomaly mediation" is an effect of integrating out the non-propagating degrees of freedom of the super-gravity multiplet, it should also by default be accounted for in the parameterisation of the soft terms.

\par Of course a more complete picture will have some drawbacks that should be overcome.  A natural model should have 3rd generation squarks lighter than the 1st and 2nd (perhaps from spatially localising the fields away from the source of supersymmetry breaking). Yet, it should also explain the generation of the Higgs sector soft masses that allow for a solution of  $\mu/B_{\mu}$ and generate electroweak symmetry breaking (EWSB) and such problems are easier to address in the context of gravity mediation.  We have checked that having the 1st and 2nd generation in the bulk and the third generation on the boundary does not effect our results, essentially as the modification of the RGEs between each case only effects the Yukawa terms and not the terms proportional to gauge couplings and in particular the dominant effect is from the gluino soft mass.

\subsection{Gauge mediation}
\par We may also expect a gauge mediated scenario compatible with this setup. In this case:
\begin{itemize}
\item The gravitino is the LSP with sneutrino or neutralino NLSP.
\item  We expect approximately flavour diagonal (if not flavour universal) soft terms.
\item The gaugino mass is $M_{3}=\frac{g^2_3}{16\pi^2}\Lambda_{\tilde{f}}$ and is not directly related to $m_{3/2}\sim \frac{F}{\sqrt{3}M_{Pl}}$,  Although we could take $\Lambda_{\tilde{f}}=\frac{F}{M_{mess}}$ and $M_{mess}\sim M_{unification}$, where $M_{unification}\sim  O(10-100)\times 1/R$ i.e. ten times the compacitifaction radius, as can be seen in figure \ref{fig:alphas5D}.
\end{itemize}
 Again the $\mu/B_{\mu}$ problem should be addressed and indeed the issue of a natural spectrum in the squark sector (light stops).  A $\mu$-term of a few hundred GeV should also lead to light Higgsinos, observable at the ILC.

\par In either scenario, we intend for naturally light stops, as can be accommodated by the large $A_t$ term, but for which we do not yet specify a fully complete picture. This setup may also be compatible with other models of supersymmetry breaking, although a ``natural spectrum'' is possible in some scenarios, light stops may not always be achievable in all models.   In the cases discussed above, the soft terms are finite and do not depend on the cutoff, all three being non-local, the first two being due to one loop diagrams that propagate in the bulk from boundary to boundary where the radius acts as a regulator on the loop diagrams.

\section{Discussion and conclusion}\label{Discussion}

\par In this paper we have explored how a five dimensional extension of the MSSM may generate a sufficiently large $A_t$ parameter to achieve the observed Higgs mass and have sub-TeV stops, perhaps observable at the LHC.  We computed the full one-loop RGEs for all supersymmetric and soft breaking parameters and then solved these equations for a given set of boundary conditions.  The results are rather interesting:  We find that Yukawa couplings may be made to unify and approximately vanish at the unification scale of the gauge couplings, for a low compactification scale, in this setup.  Further we find that the magnitude of $A_t$ follows closely that of the magnitude of the gluino mass $M_3$ and increases as the compactification scale decreases, such that a large negative $A_t$ may be achieved at low energies from a $10-10^4$ TeV compactification radius and RGE evolution from the unification scale, for a gluino mass above but not far from the current collider bounds of around $1600$ GeV.  Such a result is sufficiently general and independent of how supersymmetry is broken. A key and generic point of this work is that one may achieve larger $A_t$ terms at lower scales than are usually associated with the MSSM, by changing the UV physics and the RGEs, as such we should perhaps take the relative heavy size of the Higgs, at $125.5$ GeV as a prediction of new non-miminal physics that can effect RGEs, and not necessarily pessimistically conclude  that stops are supra-TeV in scale.  The compactification scale could be as low as a few TeV, with collider bounds on $Z'$'s being the main lower bound on this value, but electroweak precision may also be an interesting indirect constraint to explore further, due to the additional matter of this type of scenario.

\par The size of $|A_t|$ is also bounded, and cannot be too large,  as it results in an instability of the electroweak vacua to tunnel to charge and colour breaking vacua, (see for example \cite{CamargoMolina:2013sta}).   It is interesting to consider then, the relationship between gluino mass $M_3$, the radius of compactification $R$, and the magnitude of $A_t$.  For a fixed 10 TeV radius,  one cannot make $M_3$ arbitrarily large, or it induces too large an  $|A_t|$, as can be seen in figure \ref{fig:trilineargluino} and the electroweak vacuum becomes unstable.  Similarly for a fixed $M_3$, the radius cannot be made arbitrarily large, giving an indirect bound on the size of the extra dimension.

\par To extend this work, it would be interesting to explore if warped or holographic scenarios \cite{McGarrie:2010yk,Bouchart:2011va,McGarrie:2012fi} may also achieve a large $A_t$, as one expects logarithmic \cite{Pomarol:2000hp} rather than power law running  in these models. In five dimensions, one may also take advantage of non-decoupled D-terms \cite{Batra:2003nj,Batra:2004vc,Delgado:2004pr} such as in \cite{Bharucha:2013ela} to achieve a larger tree level Higgs mass.  More ambitiously, whilst in this paper these RGEs have been solved numerically at one-loop, a full and dedicated spectrum generator which implemented these 5D RGEs and various features may then give a far richer phenomenological study.


\section*{Acknowledgements}

This work is supported by the National research Foundation (South Africa) and by Campus France (Project Protea-29719RB).
AD is partially supported by Institut Universitaire de France. We also acknowledge partial support from 
the Labex-LIO (Lyon Institute of Origins) under grant ANR-10-LABX-66 and FRAMA (FR3127, F\'ed\'eration de Recherche `Andr\'e 
Marie Amp\`ere"). 


\appendix 

\section{The action and conventions}\label{appendix1}

\par In this appendix we will derive the most important RGEs for this paper.  Many of the equations are most easily computed in the superfield formalism and so we will first introduce the conventions for writing the five dimensional super Yang-Mills (SYM) action in four dimensional superspace. This action corresponds to $\mathcal{N}=2$ in the 4D perspective.  We compactify on an orbifold, $S^1/\mathbb{Z}_{2}$, such that SYM becomes a  $\mathcal{N}=1$  positive parity vector multiplet and negative parity chiral multiplet.  These conventions are based on ~\cite{Hebecker:2001ke,Mirabelli:1997aj,Cornell:2011fw}.   The maximal SYM case in five dimensions reduced to 4D superspace may be found in \cite{McGarrie:2013hca}.

\subsection{The Non-Abelian bulk action}

\par The off-shell $\mathcal{N}=1$ pure super Yang-Mills theory may be written in components:
\be
S_{5D}^{SYM}= \int d^{5} x
~\text{Tr}\left[-\frac{1}{2}(F_{MN})^2-(D_{M}\Sigma)^2-i\bar{\lambda}_{i}\gamma^M
D_{M}\gl^{i}+(X^a)^2+g_{5}\, \bar{\gl}_i[\Sigma,\gl^i]\right],
\ee
where $M,N$ run over $0,1,2,3,4$, while $\mu, \nu$ run over $0,1,2,3$. The gauge group generators and the metric are $\text{Tr}(T^{\cal A} T^{\cal B})=\frac 12 \delta^{{\cal A}{\cal B}}$ and $\eta_{MN}=\text{diag}(-1,1,1,1,1)$. The coupling $1/g^2_{5}$ has been rescaled inside the covariant derivative, $D_{M}= \partial_{M}+ig_{5} A_{M}$, where $A_{M}$ is a standard gauge vector field and $F_{MN}$ its field strength.  The other fields are a real scalar $\Sigma$, an $SU(2)_{R}$ triplet of real auxiliary fields $X^{a}$, $a=1,2,3$ and a symplectic Majorana spinor $\gl_{i}$ with $i=1,2$ which form an $SU(2)_R$ doublet. The reality condition is
\be
\gl^i= \epsilon^{ij} C\bar{\gl}_{j}^{T} 
\label{real2}
\ee
where $\epsilon^{12}=1$ and $C$ is the 5D charge conjugation matrix $C\gamma^M C^{-1}=(\gamma^M)^T$. An explicit realisation of the
Clifford algebra $\{\gamma^M,\gamma^N\}=-2\eta^{MN}$ is 
\be
\gamma^M=\left(\,\left(\begin{array}{cc}0&\sigma^\mu_{\alpha \dot{\alpha}}\\ 
\bar{\sigma}^{\mu \dot{\alpha} \alpha }&0
\end{array}\right),
\left(\begin{array}{cc}-i&0\\ 0&i\end{array}\right)\,
\right)\,,~~\mbox{and}~~~
C=\left(\begin{array}{cc}
-\epsilon_{\alpha\beta} & 0\\ 
0 & \epsilon^{\dot\alpha \dot\beta}
\end{array}\right)\,,
\ee
where $\sigma^\mu_{\alpha \dot{\alpha}}=(1,\vec{\sigma})$ and $\bar{\sigma}^{\mu \dot{\alpha} \alpha}=(1,-\vec{\sigma})$. $\alpha,\dot{\alpha}$ are spinor indices of $\text{SL}(2,C)$. For the $SU(2)_{R}$ indices we define
\be
\epsilon_{i j}=\left(\begin{array}{cc} 0&-1 \\ 1&0,
\end{array}\right) \ \ \ \ \epsilon^{ ij }=\left(\begin{array}{cc} 0&1 \\- 1&0,
\end{array}\right)
\ee
The  superalgebra is given by 
\be
\{Q^i,\bar{Q}^j\}=2\gamma^M P_M \delta^{i,j}.
\ee

\par The symplectic Majorana spinor supersymmetry parameter is $\bar{\epsilon}_i=\epsilon_{i}^{\dagger}\gamma^{0}$, which are also symplectic Majorana. To clarify notation we temporarily display all labels, writing the Dirac spinor in two component form ${\psi}^{i \, T}= (\psi_{\alpha}^{L i}, \bar{\psi}^{R \dot{\alpha} i})$ and $\bar{\psi}_i= (\psi^{R \alpha}_{i}, \bar{\psi}^L_{ \dot{\alpha} i})$.  The bar on the two component spinor denotes the complex conjugate representation of $SL(2,C)$. In particular, the reality condition~\eqref{real2} implies that
\be
\lambda^1 = \left(\begin{array}{c}
\lambda_{L \alpha}\\ 
\bar\lambda_{R}^{\dot\alpha}
\end{array}\right)~,~~~
\lambda^2 = \left(\begin{array}{c}
\lambda_{R \alpha}\\ 
-\bar\lambda_{L}^{\dot\alpha}
\end{array}\right)~,~~~
(\bar\lambda_1)^{T} = \left(\begin{array}{c}
\lambda_{R}^{\alpha}\\ 
\bar\lambda_{L \dot\alpha}
\end{array}\right)~,~~~
(\bar\lambda_2)^{T} = \left(\begin{array}{c}
-\lambda_{L}^{\alpha}\\ 
\bar\lambda_{R \dot\alpha}
\end{array}\right)~,
\ee
so the $SU(2)_{R}$ index on a two component spinor is a redundant label.

\par Next, using an orbifold $S^{1}\!/\mathbb{Z}_{2}$ the boundaries will preserve only half of the $\mathcal{N}=2$ symmetries. We choose to preserve $\epsilon_{L}$ and set $\epsilon_{R}=0$. The conjugate representations are constrained by the reality condition \ref{real2}.

\par We may therefore write a $5D$ $\mathcal{N}=1$ vector multiplet as a 4D vector multiplet and a
chiral superfield:
\begin{alignat}{1}
V=&- \theta\sigma^{\mu}\bar{\theta}A_{\mu}+i\bar{\theta}^{2}\theta\gl-
i\theta^{2}\bar{\theta}\bar{\gl}+\frac{1}{2}\bar{\theta}^{2}\theta^{2}D,\\
\Phi=& \frac{1}{\sqrt{2}}(\Sigma + i A_{5})+
\sqrt{2}\theta \chi + \theta^{2}F\,,
\label{fields2}
\end{alignat}
where the identifications between 5D and 4D fields are
\begin{equation}
D=(X^{3}-D_{5}\Sigma) \quad F=(X^{1}+iX^{2})\,,
\end{equation}
and we used $\lambda$ and $\chi$ to indicate $\lambda_{L}$ and
$-i\sqrt{2}\lambda_R$ respectively. The non-Abelian bulk action in ${\cal N}=1$  4D formalism is
\begin{equation}
S^{SYM}_{5}= \int d^{5}x \left\{\frac{1}{2}\text{Tr} \left[ \int d^{2}\theta
 W^{\alpha}W_{\alpha}+
\int d^{2}\bar{\theta}  \bar{W}_{\dot{\alpha}}\bar{W}^{\dot{\alpha}}\right]  
+ \frac{1}{2 g_{5}^2} \int d^{4}\theta  
\text{Tr}\left[e^{-2g_{5}V}\nabla_5 e^{2g_{5}V}\right]^2\right\}\, .
\label{fullaction}
\end{equation}
 $\nabla_5$ is a ``covariant'' derivative with the respect to the field $\Phi$~\cite{Hebecker:2001ke}:
\be \nabla_5 e^{2g_{5}V}=\partial_5
e^{2g_{5}V} - g_{5}\Phi^\dagger e^{2g_{5}V} - g_{5} e^{2g_{5} V} \Phi . 
\ee
Let us now focus on 5D hypermultiplets. The bulk supersymmetric action is 
\bea
S_{5D}^{H} &=& \int d^5 x[-(D_MH)^\dagger_i(D^MH^i)-i\bar{\psi}\gamma^MD_M\psi+ F^{\dagger i}F_i-g_5
\bar{\psi}\Sigma\psi+g_5 H^\dagger_i(\sigma^aX^a)^i_jH^j \nonumber\\
&& +g_5^2 H^\dagger_i\Sigma^2H^i+ig_5\sqrt{2}\bar{\psi}
\lambda^i\epsilon_{ij} H^j-i\sqrt{2}g_5H^{\dagger}_{i}
\epsilon^{ij}\bar{\gl}_{j}\psi\,].\label{hyperaction3}
\eea
$H_{i}$ are an $SU(2)_{R}$ doublet of scalars. $\psi$ is a Dirac fermion and $F_{i}$ are a doublet of scalars. With our conventions the dimensions of ($H_{i},\psi,F_{i}$) are ($\frac{3}{2},2,\frac{5}{2}$). In general the hypermultiplet matter will be in a representation of the gauge group with Dynkin index defined by $d\delta^{ab}=\text{Tr}[T^a T^b]$.

\par In the 4D superfield formulation, we again use the parity of the $P\psi_{L}=+\psi_{L}$ and $P\psi_{R}=-\psi_{R}$ to group the SUSY transformations into a positive and negative parity chiral superfields, $PH=+H$ and $PH^c=-H^c$:
\bea
H &=& H^1+\sqrt{2}\theta\psi_L+\theta^2(F_1+D_5H_{2}-g_5\Sigma H_2)\\
H^c &=& H^\dagger_2+\sqrt{2}\theta\psi_R+\theta^2(-F^{\dagger}_{2}-D_5
H^\dagger_1- g_5 H^\dagger_1\Sigma)\,.
\eea
The gauge transformations are $H\to e^{-\Lambda}H$ and $H^c\to H^ce^\Lambda$. The $\mathcal{N}=1$ action in 4D language is
\be
S_{5d}^H=\int d^5 x (\int d^4\theta [ H^\dagger e^{2g_5 V}H+H^c e^{-2g_5 V}H^{c\dagger}] + \int d^2 \theta  H^c\nabla_5 H+ \int d^2\bar{\theta}H^{c\dagger}\nabla_5 H^\dagger ) \,.
\ee


\section{Renormalisation group equations for 5D MSSM}\label{RGES5D}

\par In this section we supply the beta functions used in the main paper. We define $t=\log(Q^2/Q_0^2)$  where we take the reference scale $Q^2_0=m^2_Z$ and $\beta_A=16\pi^2 d A/dt$. For reference the gauge theory and the Higgs are in the bulk and matter fields are all localised to a brane.

\subsection{Gauge couplings}

\par The one loop beta function for the gauge couplings if $t>\text{Log}[1/R]/\text{Log}[10]$  are given by
\be
16\pi^2  \frac{d g_i[t] }{dt}= b_{MSSM}^i g^3_i[t] + b^i_{5D}g^3_i[t](S[t]-1),
\ee
where $i=1,2,3$ and $S[p]=(m_Z R) e^{p}$, where $p=t \log [10]-\log[m_Z]$. For the 4D MSSM $b^i=(33/5,1,-3)$ and for five dimensions $b^i_{5D}=(6/5,-2,-6)$. The fine structure constants may be defined from $\alpha_i= g^2_i/4\pi$. Instead one could consider including one Kaluza-Klein mode at a time, in which case one finds 
\be
\beta_{g_i}=\frac{g^3_{i}}{16\pi^2}\left[b^i_{MSSM}+n \tilde{b}^i_{5D} \right] \ \  \   ,  \ \   \  \beta_{M_i}=\frac{2 g^2_{i}M_i}{16\pi^2}\left[b^i_{MSSM}+ n \tilde{b}^i_{5D} \right].
\ee
We instead use the Kaluza-Klein summed expression above.

\subsection{Yukawa couplings}

\par
\begin{figure}[htp!]
\begin{center}
\includegraphics[scale=0.45]{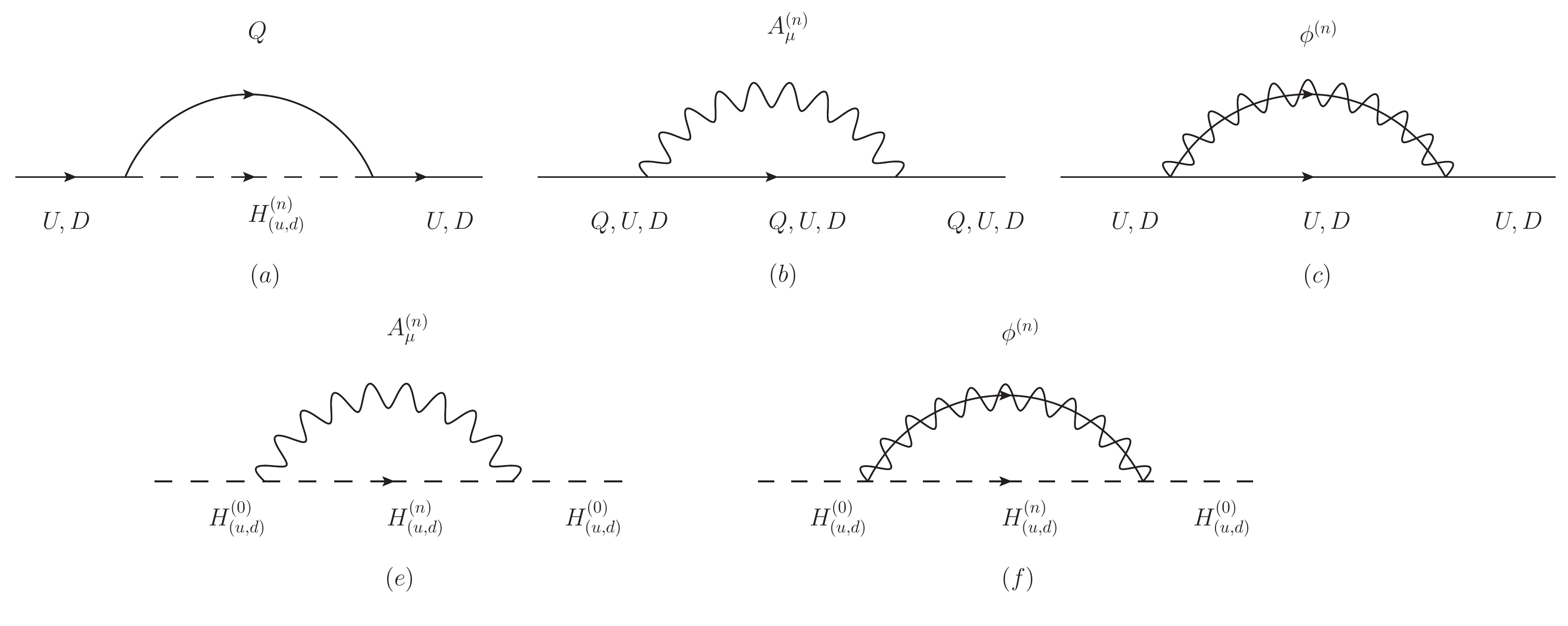}
\caption{{\it The wavefunction renormalisation contribution for the five dimensional Yukawas.}} 
\label{fig:Yukawas}
\end{center}
\end{figure}

 The beta functions for the Yukawa couplings may be related to the matrices of anomalous dimensions
\be
\beta_Y^{ijk}=\gamma^i_n Y^{njk}+\gamma^i_n Y^{ink}+\gamma^k_n Y^{ijn}.
\ee
The one-loop RGEs for Yukawa couplings in the 4D MSSM are given by (see figure \ref{fig:Yukawas})
\begin{align} 
\beta_{Y_u}^{(1)} & =  
3 {Y_u  Y_{u}^{\dagger}  Y_u} + {Y_u  Y_{d}^{\dagger}  Y_d} -\frac{1}{15} Y_u \Big(13 g_{1}^{2}  + 45 g_{2}^{2}  + 80 g_{3}^{2}-45 \mbox{Tr}\Big({Y_u  Y_{u}^{\dagger}}\Big)   \Big) \\ 
\beta_{Y_d}^{(1)} & =  
3 {Y_d  Y_{d}^{\dagger}  Y_d}  + {Y_d  Y_{u}^{\dagger}  Y_u} + Y_d \Big(-3 g_{2}^{2}   -\frac{16}{3} g_{3}^{2}  -\frac{7}{15} g_{1}^{2}  + \mbox{Tr}\Big({Y_e  Y_{e}^{\dagger}}\Big)+ 3 \mbox{Tr}\Big({Y_d  Y_{d}^{\dagger}}\Big) \Big)\\ 
\beta_{Y_e}^{(1)} & =  
3 {Y_e  Y_{e}^{\dagger}  Y_e}  + Y_e \Big(-3 g_{2}^{2}   -\frac{9}{5} g_{1}^{2}  + \mbox{Tr}\Big({Y_e  Y_{e}^{\dagger}}\Big)+ 3 \mbox{Tr}\Big({Y_d  Y_{d}^{\dagger}}\Big) \Big).
\end{align}

\par The five dimensional contribution is given by
\bea
\beta^{(1)}_{(5D)Y_u}[t] &=& Y_u \left[ \left(6Y^{\dagger}_u Y_u + 2Y^{\dagger}_d Y_d \right)-  \left(\frac{34}{30}g_1^2+\frac{9}{2}g_2^2+\frac{32}{3}g_3^2\right) \right] \\
\beta^{(1)}_{(5D)Y_d}[t] &=&  Y_d \left[ (6Y^{\dagger}_d Y_d + 2Y^{\dagger}_u Y_u) -  \left(\frac{19}{30}g_1^2+\frac{9}{2}g_2^2+\frac{32}{3}g_3^2\right) \right] \\
\beta^{(1)}_{(5D)Y_e}[t] &=&  Y_e \left[ 6Y^{\dagger}_e Y_e -  \left(\frac{33}{10}g_1^2+\frac{9}{2}g_2^2\right) \right].
\eea
\subsection{Trilinear soft breaking parameters}
\par
\begin{figure}[htb!]
\begin{center}
\includegraphics[scale=0.5]{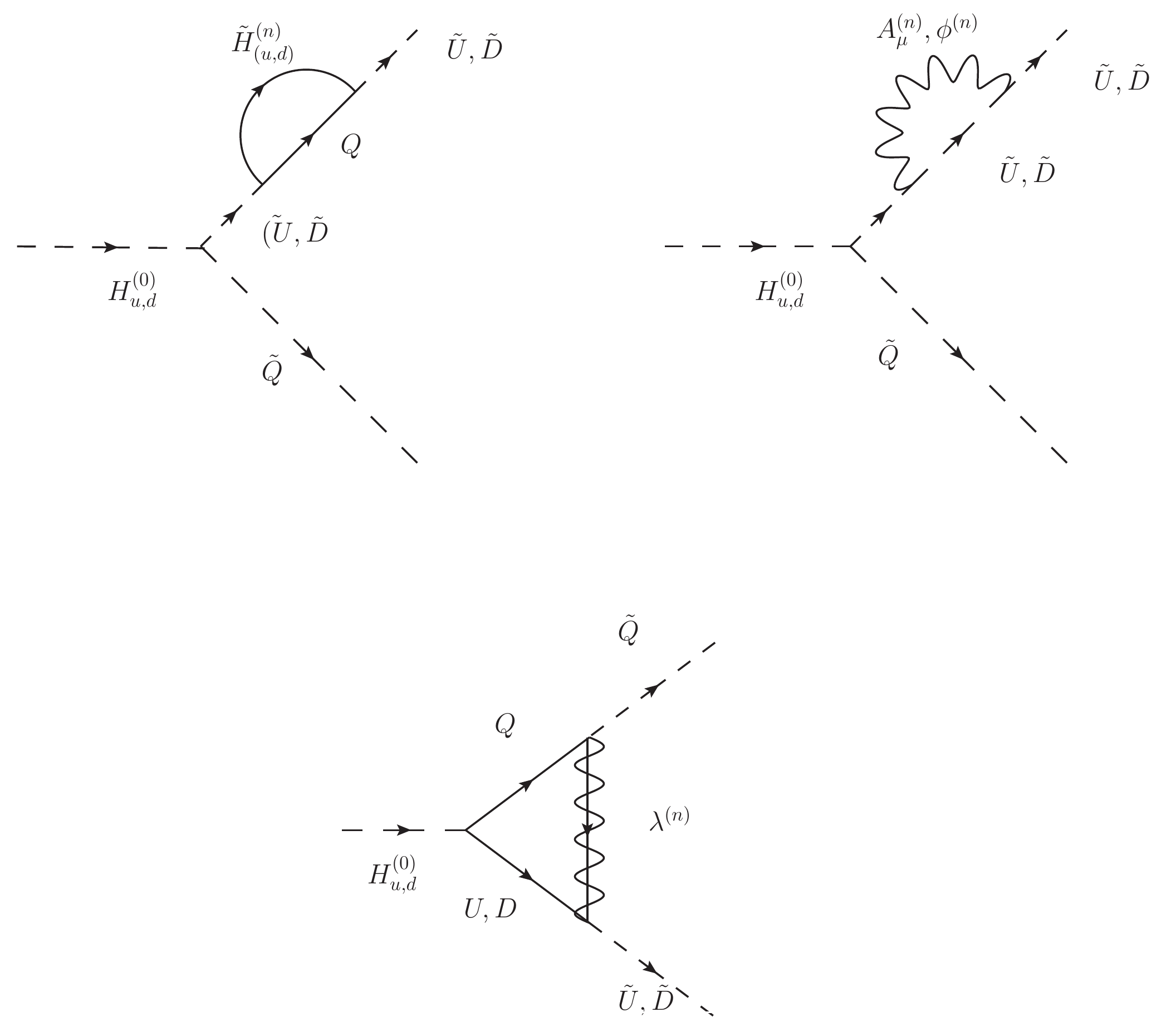}
\caption{{\it The diagrams contributing to the five dimensional RGEs of the Trilinear soft breaking parameters.}} 
\label{fig:Trilinears}
\end{center}
\end{figure}

 The 4D MSSM soft breaking parameters at one loop, as pictured in figure \ref{fig:Trilinears} in are given by
\bea
\beta_{A_u}^{(1)}  &=&  
+2 {Y_u Y_{d}^{\dagger} A_d} +4 {Y_u Y_{u}^{\dagger} A_u} +{A_u Y_{d}^{\dagger} Y_d}+5 {A_u Y_{u}^{\dagger} Y_u} -\frac{13}{15} g_{1}^{2} A_u -3 g_{2}^{2} A_u -\frac{16}{3} g_{3}^{2} A_u \nonumber \\ 
 &&+3 A_u \mbox{Tr}\Big({Y_u Y_{u}^{\dagger}}\Big) +Y_u \Big(6 g_{2}^{2} M_2 + 6 \mbox{Tr}\Big({Y_{u}^{\dagger} A_u}\Big) + \frac{26}{15} g_{1}^{2} M_1 + \frac{32}{3} g_{3}^{2} M_3 \Big)\\ 
\beta_{A_d}^{(1)} &=&  
+4 {Y_d Y_{d}^{\dagger} A_d} +2 {Y_d Y_{u}^{\dagger} A_u} +5 {A_d Y_{d}^{\dagger} Y_d} +{A_d Y_{u}^{\dagger} Y_u}-\frac{7}{15} g_{1}^{2} A_d -3 g_{2}^{2} A_d -\frac{16}{3} g_{3}^{2} A_d \nonumber \\ 
 &&+3 A_d \mbox{Tr}\Big({Y_d Y_{d}^{\dagger}}\Big) +A_d \mbox{Tr}\Big({Y_e Y_{e}^{\dagger}}\Big) +Y_d \Big(2 \mbox{Tr}\Big({Y_{e}^{\dagger} A_e}\Big) \nonumber \\ 
&& + 6 g_{2}^{2} M_2 + 6 \mbox{Tr}\Big({Y_{d}^{\dagger}  A_d}\Big) + \frac{14}{15} g_{1}^{2} M_1 + \frac{32}{3} g_{3}^{2} M_3 \Big)\\ 
\beta_{A_e}^{(1)} &=&  
+4 {Y_e Y_{e}^{\dagger} A_e} +5 {A_e Y_{e}^{\dagger} Y_e} -\frac{9}{5} g_{1}^{2} A_e -3 g_{2}^{2} A_e +3 A_e \mbox{Tr}\Big({Y_d Y_{d}^{\dagger}}\Big) +A_e \mbox{Tr}\Big({Y_e Y_{e}^{\dagger}}\Big) \nonumber \\ 
 &&+Y_e \Big(2 \mbox{Tr}\Big({Y_{e}^{\dagger} A_e}\Big) + 6 g_{2}^{2} M_2 + 6 \mbox{Tr}\Big({Y_{d}^{\dagger} A_d}\Big) + \frac{18}{5} g_{1}^{2} M_1 \Big).
\eea
\par In the 5D MSSM these are given by:
\begin{eqnarray}
\beta^{(1)}_{(5D) A_u}[t] &=& A_u \left(\left(18Y^{\dagger}_u Y_u + 2Y^{\dagger}_d Y_d \right)- \left(\frac{34}{30}g_1^2+\frac{9}{2}g_2^2+\frac{32}{3}g_3^2\right) \right)+ 4A_d Y^{\dagger}_d Y_u \nonumber\\
&&+ Y_u \left(\frac{34}{15}g_1^2 M_1 + 9g_2^2 M_2 +\frac{64}{3}g_3^2 M_3 \right)\\
\beta^{(1)}_{(5D) A_d}[t] &=& A_d \left(\left(18Y^{\dagger}_d Y_d + 2Y^{\dagger}_u Y_u \right)-\left(\frac{19}{30}g_1^2+\frac{9}{2}g_2^2+\frac{32}{3}g_3^2\right)\right)\nonumber\\
 &&+ 4A_u Y^{\dagger}_u Y_d  + 2A_e Y^{\dagger}_e Y_d + Y_d \left[\frac{19}{15}g_1^2 M_1 + 9g_2^2 M_2 +\frac{64}{3}g_3^2 M_3 \right]\\
\beta^{(1)}_{(5D) A_e}[t] &=& A_e \left( 18Y^{\dagger}_e Y_e - \left(\frac{33}{10}g_1^2+\frac{9}{2}g_2^2\right) \right) + 6A_d Y^{\dagger}_d Y_e + Y_e \left(\frac{33}{5}g_1^2 M_1 + 9g_2^2 M_2 \right). \nonumber \\
\end{eqnarray}
\subsection{Soft masses}
\par
\begin{figure}[htp!]
\begin{center}
\includegraphics[scale=0.6]{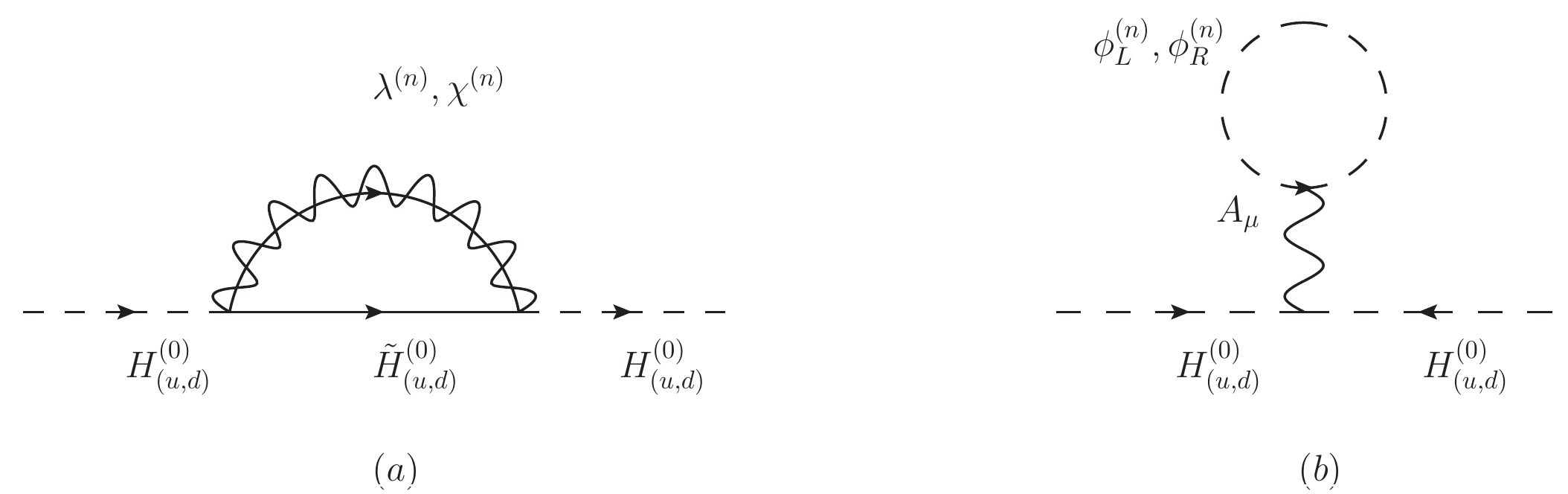}
\caption{{\it The diagrams for the five dimensional renormalisation group equations of the soft scalar masses at one loop.}} 
\label{fig:softscalarmasses}
\end{center}
\end{figure}

 We expect the gaugino soft masses to run following
\be
\beta^{(1)}_{M_i}[t]= 2b_{MSSM}^i M_i[t]g^2_i[t] + 2b^i_{(5D)}M_i[t]g^2_i[t] (S[t]-1).
\ee
The scalar soft masses have five dimensional RGE contributions as pictured in figure \ref{fig:softscalarmasses}.  The four dimensional MSSM  contribution is
\bea
\beta_{m_q^2}^{(1)} & = & 
-\frac{2}{15} g_{1}^{2} {\bf 1} |M_1|^2 -\frac{32}{3} g_{3}^{2} {\bf 1} |M_3|^2 -6 g_{2}^{2} {\bf 1} |M_2|^2 +2 m_{H_d}^2 {Y_{d}^{\dagger} Y_d} +2 m_{H_u}^2 {Y_{u}^{\dagger} Y_u} +2 {A_{d}^{\dagger} A_d} \nonumber \\ 
& &+2 {A_{u}^{\dagger} A_u} +{m_q^2 Y_{d}^{\dagger} Y_d}+{m_q^2 Y_{u}^{\dagger} Y_u}+2 {Y_{d}^{\dagger} m_d^2 Y_d} +{Y_{d}^{\dagger}  Y_d  m_q^2}+2 {Y_{u}^{\dagger} m_u^2 Y_u} \nonumber \\ 
 &&+{Y_{u}^{\dagger} Y_u  m_q^2}+\frac{1}{\sqrt{15}} g_1 {\bf 1} \sigma_{1,1} \\ 
\beta_{m_u^2}^{(1)} & = & 
-\frac{32}{15} g_{1}^{2} {\bf 1} |M_1|^2 -\frac{32}{3} g_{3}^{2} {\bf 1} |M_3|^2 +4 m_{H_u}^2 {Y_u Y_{u}^{\dagger}} +4 {A_u A_{u}^{\dagger}} +2 {m_u^2 Y_u Y_{u}^{\dagger}} +4 {Y_u m_q^2 Y_{u}^{\dagger}} \nonumber \\ 
& &+2 {Y_u Y_{u}^{\dagger} m_u^2} -4 \frac{1}{\sqrt{15}} g_1 {\bf 1} \sigma_{1,1} \\ 
\beta_{m_d^2}^{(1)} & =  &
-\frac{8}{15} g_{1}^{2} {\bf 1} |M_1|^2 -\frac{32}{3} g_{3}^{2} {\bf 1} |M_3|^2 +4 m_{H_d}^2 {Y_d Y_{d}^{\dagger}} +4 {A_d A_{d}^{\dagger}} +2 {m_d^2 Y_d Y_{d}^{\dagger}} +4 {Y_d m_q^2 Y_{d}^{\dagger}} \nonumber \\ 
& &+2 {Y_d Y_{d}^{\dagger} m_d^2} +2 \frac{1}{\sqrt{15}} g_1 {\bf 1} \sigma_{1,1} \\ 
\beta_{m_l^2}^{(1)} & = & 
-\frac{6}{5} g_{1}^{2} {\bf 1} |M_1|^2 -6 g_{2}^{2} {\bf 1} |M_2|^2 +2 m_{H_d}^2 {Y_{e}^{\dagger} Y_e} +2 {A_{e}^{\dagger} A_e} +{m_l^2 Y_{e}^{\dagger} Y_e}+2 {Y_{e}^{\dagger} m_e^2 Y_e} \nonumber \\ 
& &+{Y_{e}^{\dagger} Y_e m_l^2}- \sqrt{\frac{3}{5}} g_1 {\bf 1} \sigma_{1,1} \\ 
\beta_{m_e^2}^{(1)} & =  &
-\frac{24}{5} g_{1}^{2} {\bf 1} |M_1|^2 +2 \Big(2 m_{H_d}^2 {Y_e Y_{e}^{\dagger}} + 2 {A_e  A_{e}^{\dagger}} + 2 {Y_e m_l^2 Y_{e}^{\dagger}}  + {m_e^2  Y_e  Y_{e}^{\dagger}} + {Y_e  Y_{e}^{\dagger}  m_e^2}\Big)\nonumber \\ 
& &+2 \sqrt{\frac{3}{5}} g_1 {\bf 1} \sigma_{1,1} 
\eea 
where 
\begin{align} 
\sigma_{1,1} & = \sqrt{\frac{3}{5}} g_1 \Big(-2 \mbox{Tr}\Big({m_u^2}\Big) - \mbox{Tr}\Big({m_l^2}\Big) - m_{H_d}^2 + m_{H_u}^2 + \mbox{Tr}\Big({m_d^2}\Big) + \mbox{Tr}\Big({m_e^2}\Big) + \mbox{Tr}\Big({m_q^2}\Big)\Big).
\end{align} 

\par In the 5D MSSM these are given by:
\bea
\beta_{(5D) m_q^2}^{(1)} & = &\left[ 
-\frac{4}{15} g_{1}^{2} {\bf 1} |M_1|^2 -\frac{64}{3} g_{3}^{2} {\bf 1} |M_3|^2 -9 g_{2}^{2} {\bf 1} |M_2|^2 +\frac{\sqrt{2}}{\sqrt{15}} g_1 {\bf 1} \sigma_{1,1}\right]\\
\beta_{(5D) m_l^2}^{(1)} & =  &
\left[-\frac{12}{5} g_{1}^{2} {\bf 1} |M_1|^2 -9 g_{2}^{2} {\bf 1} |M_2|^2 - \sqrt{\frac{6}{5}} g_1 {\bf 1} \sigma_{1,1} \right]\\
\beta_{(5D) m_u^2}^{(1)} & =  &
\left[-\frac{64}{15} g_{1}^{2} {\bf 1} |M_1|^2 -\frac{64}{3} g_{3}^{2} {\bf 1} |M_3|^2 -4 \frac{\sqrt{2}}{\sqrt{15}} g_1 {\bf 1} \sigma_{1,1} \right]\\
\beta_{(5D) m_d^2}^{(1)} & =  &
\left[-\frac{16}{15} g_{1}^{2} {\bf 1} |M_1|^2 -\frac{64}{3} g_{3}^{2} {\bf 1} |M_3|^2 +2 \frac{\sqrt{2}}{\sqrt{15}} g_1 {\bf 1} \sigma_{1,1}  \right]\\
\beta_{(5D) m_e^2}^{(1)} & =  &
\left[-\frac{48}{5} g_{1}^{2} {\bf 1} |M_1|^2 +2 \sqrt{\frac{6}{5}} g_1 {\bf 1} \sigma_{1,1} \right].
\eea
The one-loop RGE's  for the two Higgs doublet soft masses in the  4D MSSM are given by
\bea
\beta_{m_{H_d}^2}^{(1)} & =  &
-\frac{6}{5} g_{1}^{2} |M_1|^2 -6 g_{2}^{2} |M_2|^2 - \sqrt{\frac{3}{5}} g_1 \sigma_{1,1} +6 m_{H_d}^2 \mbox{Tr}\Big({Y_d Y_{d}^{\dagger}}\Big)+2 m_{H_d}^2 \mbox{Tr}\Big({Y_e Y_{e}^{\dagger}}\Big)\nonumber\\
& &+6 \mbox{Tr}\Big({A_d^* A_{d}^{T}}\Big)+2 \mbox{Tr}\Big({A_e^* A_{e}^{T}}\Big)+6 \mbox{Tr}\Big({m_d^2 Y_d Y_{d}^{\dagger}}\Big)+2 \mbox{Tr}\Big({m_e^2 Y_e Y_{e}^{\dagger}}\Big)\nonumber\\
& &+2 \mbox{Tr}\Big({m_l^2 Y_{e}^{\dagger} Y_e}\Big) +6 \mbox{Tr}\Big({m_q^2 Y_{d}^{\dagger} Y_d}\Big) \\ 
\beta_{m_{H_u}^2}^{(1)} & =  &
-\frac{6}{5} g_{1}^{2} |M_1|^2 -6 g_{2}^{2} |M_2|^2 +\sqrt{\frac{3}{5}} g_1 \sigma_{1,1} +6 m_{H_u}^2 \mbox{Tr}\Big({Y_u Y_{u}^{\dagger}}\Big) \nonumber\\
&&+6 \mbox{Tr}\Big({A_u^* A_{u}^{T}}\Big) +6 \mbox{Tr}\Big({m_q^2 Y_{u}^{\dagger} Y_u}\Big) +6 \mbox{Tr}\Big({m_u^2 Y_u Y_{u}^{\dagger}}\Big).
\eea
In 5D MSSM the two Higgs doublet soft masses obey the RGE's 
\begin{align} 
\beta_{(5D) m_{H_d}^2}^{(1)} & =  
\left[-\frac{12}{5} g_{1}^{2} |M_1|^2 -9 g_{2}^{2} |M_2|^2 - 2\sqrt{\frac{3}{5}} g_1 \sigma_{1,1}\right]\\ 
\beta_{(5D) m_{H_u}^2}^{(1)} & =  
 \left[-\frac{12}{5} g_{1}^{2} |M_1|^2 -9 g_{2}^{2} |M_2|^2 +2\sqrt{\frac{3}{5}} g_1 \sigma_{1,1}\right].
\end{align}

\subsection{Bilinear parameters $\mu$ and $B_{\mu}$}

\par The one-loop beta function of $\mu$ and $B_{\mu}$ in the 4D MSSM are given by:
\bea 
\beta_{\mu}^{(1)} & =  &
3 \mu \mbox{Tr}\Big({Y_d Y_{d}^{\dagger}}\Big) -\frac{3}{5} \mu \Big(5 g_{2}^{2} -5 \mbox{Tr}\Big({Y_u Y_{u}^{\dagger}}\Big) + g_{1}^{2}\Big) + \mu \mbox{Tr}\Big({Y_e Y_{e}^{\dagger}}\Big) \\
\beta^{(1)}_{B_{\mu}} & =  &
3 B_{\mu} \mbox{Tr}\Big({Y_d Y_{d}^{\dagger}}\Big) -\frac{3}{5} B_{\mu} \Big(5 g_{2}^{2} -5 \mbox{Tr}\Big({Y_u Y_{u}^{\dagger}}\Big) + g_{1}^{2}\Big) + B_{\mu} \mbox{Tr}\Big({Y_e Y_{e}^{\dagger}}\Big) \nonumber\\
&&+6 \mu \mbox{Tr}\Big({A_d Y_{d}^{\dagger}}\Big) +\frac{6}{5} \mu \Big(5 g_{2}^{2} M_2 +5 \mbox{Tr}\Big({A_u Y_{u}^{\dagger}}\Big) + g_{1}^{2} M_1 \Big) + 2\mu \mbox{Tr}\Big({A_e Y_{e}^{\dagger}}\Big).
\eea 
In the 5D MSSM these are given by:
\bea 
\beta_{(5D)\mu}^{(1)} & = & \mu \left[-\frac{6}{5} g_{1}^{2} -\frac{9}{2} g_{2}^{2} \right] \\
\beta^{(1)}_{B_{\mu}} & =  &
 - B_{\mu} \Big(\frac{9}{2} g_{2}^{2} + \frac{6}{5} g_{1}^{2}\Big) + \mu \Big(9 g_{2}^{2} M_2+\frac{12}{5}g_{1}^{2} M_1\Big) .
\eea 


\bibliographystyle{JHEP}
\bibliography{LargeAt}

\end{document}